\documentclass[aps,prl,notitlepage,amsmath,amssymb,amsfonts,superscriptaddress,twocolumn]{revtex4-2}

\makeatletter
\def\@hangfrom@section#1#2#3{\normalsize\@hangfrom{#1#2}#3}
\def\@hangfroms@section#1#2{\normalsize#1#2}
\makeatother

\usepackage{graphicx,soul}
\usepackage{dcolumn}
\usepackage{bm}
\usepackage{bbm} 

\usepackage{dsfont}
\usepackage{enumerate}

\usepackage[dvipsnames, x11names]{xcolor}
\usepackage[colorlinks=true]{hyperref}
\hypersetup{
    colorlinks=true,
    citecolor=blue,
    linkcolor=red,
    urlcolor=RoyalBlue,
    anchorcolor=Purple
}

\usepackage[percent]{overpic}
\usepackage{tabularx}
\usepackage{multirow}
\usepackage{CJK}
\usepackage{orcidlink}

\usepackage{physics}

\usepackage{pdfpages}
\usepackage{pgffor}
\makeatletter
\AtBeginDocument{\let\LS@rot\@undefined}
\makeatother

\newcommand{\bluefour}[1]{\textcolor{Blue4}{#1}}

\def\be{\begin{equation}}
\def\ee{\end{equation}}
\def\bea{\begin{eqnarray}}
\def\eea{\end{eqnarray}}

\begin{document}

\begin{CJK*}{UTF8}{}

    \title{
    Emergent Gauge Flux and Spin Ordering in Magnetized Triangular Spin Liquids: \\
    Applications to Hofstadter-Hubbard Model
    }

    \author{Jiahao~Yang~(\CJKfamily{gbsn}杨家豪) \orcidlink{0000-0001-7670-2218}}
    \affiliation{International Center for Quantum Materials, School of Physics, Peking University, Beijing 100871, China}
    \affiliation{Beijing Key Laboratory of Quantum Devices, Peking University, Beijing 100871, China}

    \author{Hao Tian \orcidlink{0009-0007-2238-0883}}
    \affiliation{International Center for Quantum Materials, School of Physics, Peking University, Beijing 100871, China}

    \author{Si-Yu~Pan \orcidlink{0009-0009-2706-3523}}
    \affiliation{International Center for Quantum Materials, School of Physics, Peking University, Beijing 100871, China}

    \author{Gang~v.~Chen \orcidlink{0000-0001-9339-6398}} 
    \email{chenxray@pku.edu.cn}
    \affiliation{International Center for Quantum Materials, School of Physics, Peking University, Beijing 100871, China}
    \affiliation{Beijing Key Laboratory of Quantum Devices, Peking University, Beijing 100871, China}
    \affiliation{Collaborative Innovation Center of Quantum Matter, 100871, Beijing, China}

    \begin{abstract}
    Motivated by recent progress in moir\'{e} superlattices and spin-$1/2$ triangular-lattice antiferromagnets, we study how orbital magnetic flux and Zeeman coupling compete or cooperate in generating internal U(1) gauge flux in a triangular spin liquid.
    We show that orbital flux favors a chiral spin liquid with staggered internal flux, whereas Zeeman coupling destabilizes the spinon Fermi-pocket state toward a Landau-level state with spontaneous uniform internal flux and conical spin order.
    We demonstrate this mechanism in a spin-$1/2$ $J_1$-$J_2$-$J_{\chi}$ model, and further identify thermal Hall and magnetic signatures that distinguish these regimes, with potential applications to moir\'{e} Hofstadter-Hubbard systems and triangular-lattice antiferromagnets.
    \end{abstract}

    \date{\today}
    \maketitle

\end{CJK*}


\noindent\bluefour{\it Introduction.}---The Shubnikov--de Haas and de Haas--van Alphen effects originate from Landau quantization of electronic states in a magnetic field and appear as quantum oscillations in transport and magnetization, respectively~\cite{shoenberg1984MagneticOscillationsMetals}. 
In a conventional insulator without a charge Fermi surface, such oscillations are absent. 
For weak Mott insulating spin liquids, however, 
it has been proposed that an orbital magnetic field can induce an emergent internal gauge flux for fermionic spinons, 
enabling Landau quantization of neutral spin excitations and thereby producing quantum oscillations~\cite{motrunich2006OrbitalMagneticField,sodemann2018QuantumOscillations}. 
As the applied flux varies, the induced  internal gauge flux evolves and spinon Landau levels are successively filled, and the corresponding integer-filled states provide a route to chiral spin liquid (CSL) phases. 
This appealing scenario faces two central challenges. One is that in ordinary solids the microscopic lattice spacing strongly suppresses the orbital flux through each plaquette. The other is the unavoidable Zeeman coupling, which may polarize the spin degrees of freedom.

Recent progress in moir\'{e} superlattice systems has substantially changed this situation. 
Because of the large moir\'{e} lattice constant, an appreciable orbital flux can thread each moir\'{e} unit cell, while layer or valley textures may further generate topological flat bands with nontrivial gauge flux~\cite{wu2019TopologicalInsulatorsTwisted}. 
In particular, triangular Hubbard models proposed for transition-metal dichalcogenide moir\'{e} systems naturally extend to triangular Hofstadter-Hubbard (HH) models, where Hofstadter minibands, Berry curvature, and interaction-driven Mott physics are intertwined~\cite{wu2018HubbardModel,zang2021HartreeFockStudy}. 
These systems provide an experimentally relevant setting for studying the interplay between quantum Hall physics and magnetically ordered Mott phases, and have motivated extensive work on chiral spin liquids~\cite{kuhlenkamp2024AspectsProbes,kuhlenkamp2024ChiralPseudospin,divic2024ChiralSpin,gallegos2025QuantumHall,chen2016SyntheticgaugefieldStabilization,zhang2026PhasesCriticality,yao2021TopologicalChiral,yao2026SurfaceChiral} and superconductivity upon doping~\cite{divic2025AnyonSuperconductivityTopological,chen2026TopologicalChiral,kuhlenkamp2025RobustSuperconductivity,pichler2025MicroscopicMechanism}. 
In most existing studies, however, the Zeeman coupling is either bypassed by using valley or layer degrees of freedom as pseudospin or is simply neglected~\cite{kuhlenkamp2024ChiralPseudospin}. 
Recent numerical works have begun to address Zeeman-field effects in the triangular spin-liquid regime~\cite{keselman2025J1J2Triangular,bader2025J1J2Triangular}, but a clear analytic and physical understanding of how the Zeeman coupling competes or cooperates with orbital-flux-induced gauge physics is still lacking.

This question is also relevant to spin-$1/2$ triangular-lattice antiferromagnets, including rare-earth compounds~\cite{shen2016EvidenceSpinon,shen2018FractionalizedExcitations,bag2024EvidenceDiracQuantum,wu2025SpinDynamics} and cobalt-based materials~\cite{lee2021TemporalField,zhong2019StrongQuantum,zhong2020FrustratedMagnetism,zhu2024ContinuumExcitations}, where proximate spin-liquid behavior and unusual magnetic-field responses have been widely discussed. 
To capture both moir\'{e} HH systems and triangular antiferromagnets within a unified framework, we study a spin-$1/2$ $J_1$-$J_2$-$J_\chi$ model in a Zeeman field. 
Here the scalar spin-chirality term $J_\chi$ encodes the orbital-flux effect of the HH model in the Mott regime, while the Zeeman term directly induces finite magnetization~\cite{SuppMat}. 
This setup allows us to address the central issue of how orbital flux and Zeeman coupling generate distinct internal gauge-field responses in a spin-liquid background, and how these responses lead to discernible experimental signatures.

Our results reveal this intertwined effect explicitly.
Orbital flux favors a staggered internal flux and stabilizes the CSL, whereas Zeeman coupling drives the magnetized Dirac spin liquid toward a spinon Landau-level state with spontaneous uniform internal U(1) gauge flux. 
This state is energetically favored over the competing Fermi-pocket state and is tied, through the monopole condensation, to $K$-point 120$^\circ$ cone order. 
We further identify thermal Hall and spin-structure-factor signatures that distinguish these regimes and provide concrete probes for moir\'{e} Hofstadter-Hubbard systems and triangular-lattice antiferromagnets.

\begin{figure}[t]
    \centering
    \begin{overpic}
        [width=0.4\linewidth]{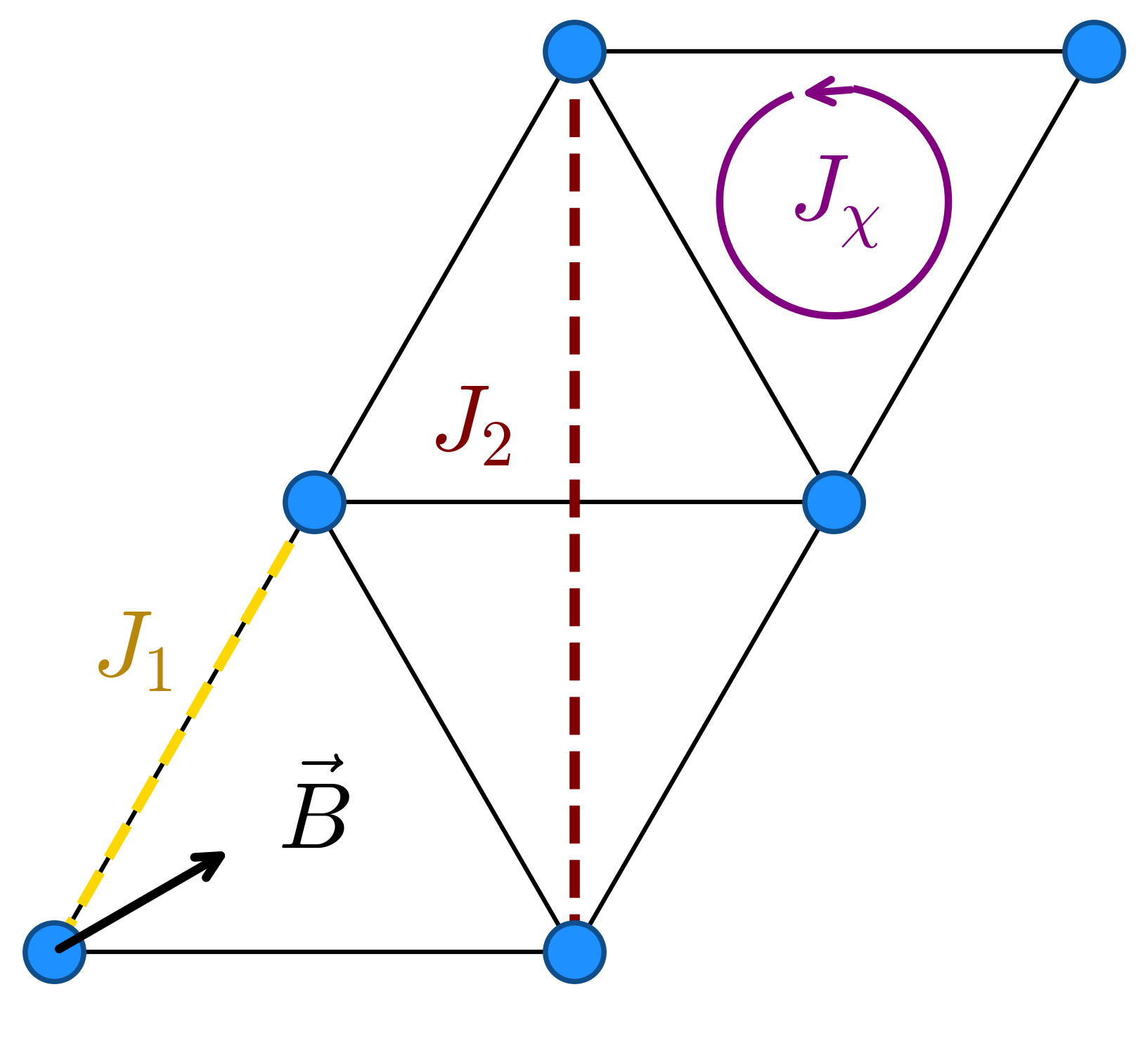}
        \put(0,65){(a)}
    \end{overpic}
    \begin{overpic}
        [width=0.58\linewidth]{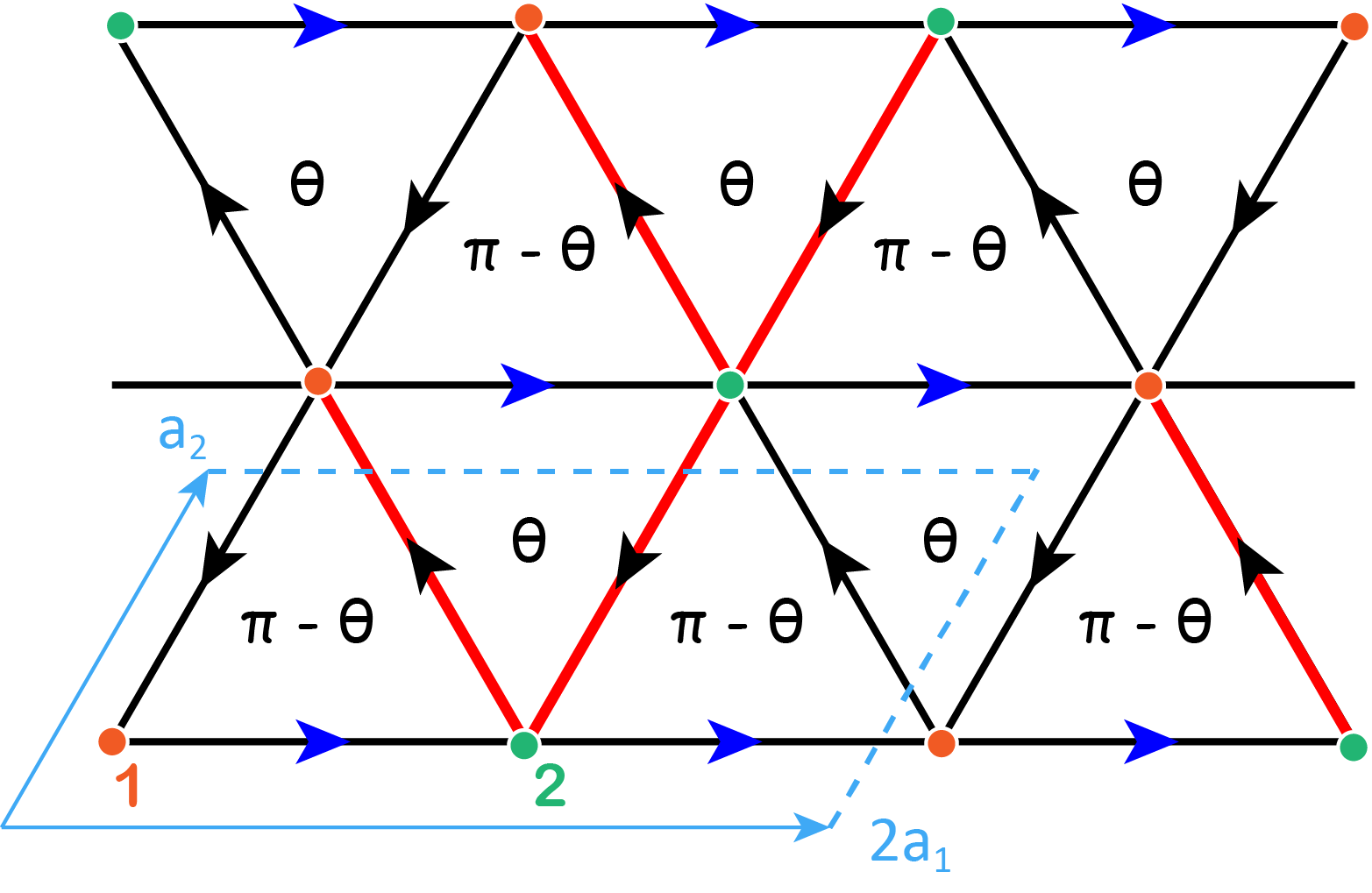}
        \put(0,46){(b)}
    \end{overpic}
    \caption{
        (a) Schematic illustration of the triangular lattice $J_1$-$J_2$-$J_\chi$ model under a magnetic field $B$. 
        (b) The $[\theta,\pi-\theta]$ state with staggered $\theta$ flux.
        The phase $e^{i\phi_{ij}}$ takes ${\phi_{ij}=\mp\theta}$ when the spinon hops 
        parallel (antiparallel) to the blue arrows, and $\phi_{ij}=0$ 
        when the spinon hops along black arrows.
        For the red lines, $\phi_{ij}$ further acquires a phase of $\pi$.
    }
    \label{fig:trian_latt}
\end{figure}

\noindent\bluefour{\it Model.}---The spin-$1/2$ $J_1$-$J_2$-$J_\chi$ model is defined by
 \begin{equation}
    \mathcal{H}=J_1\sum_{\langle ij\rangle}\bm{S}_i\cdot\bm{S}_j
    + J_2\sum_{\langle\langle ij\rangle\rangle}\bm{S}_i\cdot\bm{S}_j
    + 
    J_\chi\sum_{\vartriangle}\bm{S}_i\cdot(\bm{S}_j\times\bm{S}_k),
    \label{eq:H_J1J2Jchi}
\end{equation}
where $J_1$ and $J_2$ are antiferromagnetic first- and second-neighbor exchange couplings, and $J_\chi$ is the scalar spin-chirality interaction shown in Fig.~\ref{fig:trian_latt}(a). In the Mott regime of the HH model, this term arises from third-order perturbation theory and depends on the orbital flux $\phi_{\vartriangle}$ through a triangular plaquette as $J_\chi \propto \sin \phi_{\vartriangle}$~\cite{motrunich2006OrbitalMagneticField,SuppMat,delannoy2005NeelOrder,huang2024MagneticFieldInduced}. In strong-Mott materials, $J_\chi$ is typically small and pairwise exchanges dominate. Because the model preserves spin-rotation symmetry, the Zeeman field may be chosen along any direction; for concreteness, we take it along the $z$ axis, which gives the Zeeman term $-B\sum_i S_i^z$.

The phase diagram of this model is well established from earlier studies~\cite{gong2017GlobalPhaseDiagram,wietek2017ChiralSpinLiquid,iqbal2016SpinLiquid,hu2016VariationalMonte}. 
At $J_{\chi}=0$, the system realizes the $\sqrt{3}\times\sqrt{3}$ ordered state for $J_2/J_1 \lesssim 0.08$, the stripe state for $J_2/J_1 \gtrsim 0.15$, and a U(1) Dirac spin liquid (DSL) in the intermediate regime $0.08 \lesssim J_2/J_1 \lesssim 0.15$. 
A nonzero $J_{\chi}$ stabilizes a chiral spin liquid (CSL), identified with the Kalmeyer-Laughlin $\nu = 1/2$ bosonic fractional quantum Hall state, over a broad parameter range~\cite{gong2017GlobalPhaseDiagram,hu2016VariationalMonte}. 
The orbital flux therefore favors the formation of the CSL in this model. 
Since the ordered phases are comparatively well understood, we focus here on the magnetic response of the two spin-liquid states and on the experimental signatures that distinguish them.


\noindent\bluefour{\it Triangular Spin liquids.}---In the intermediate nonmagnetic regime, the relevant low-energy state is the U(1) DSL with gapless Dirac spinons and gapped CSL~\cite{iqbal2016SpinLiquid,hu2016VariationalMonte}.
To capture the spin liquids physics of Eq.~\eqref{eq:H_J1J2Jchi}, we introduce Abrikosov 
fermionic operators $f_{i\alpha}$, with ${\alpha = \uparrow, \downarrow}$ 
being the spin up and down index.
The spin-1/2 operators can be expressed as the fermion bilinears 
${S_i^a = \frac{1}{2} \sum_{\alpha\beta} f^\dagger_{i\alpha}\sigma^a_{\alpha\beta} f_{i\beta}^{}}$,
with Pauli matrices ${\sigma^a (a = x, y, z)}$ and a local SU(2) gauge symmetry~\cite{wen2002QuantumOrders}.
The physical Hilbert space is restored by imposing a local single-occupancy constraint, 
$\sum_{\alpha} f^\dagger_{i\alpha}f_{i\alpha}=1$.
Applying the parton mean-field decoupling, 
we obtain the spinon mean-field theory \cite{wen2007QuantumFieldTheory}, 
\begin{equation}
    \begin{aligned}
        H_{\text{MF}} & =-t\sum_{\langle {ij}\rangle,\alpha}
         \left( e^{i\phi_{{ij}}}
       f_{{i}\alpha}^\dagger f_{{j}\alpha}^{} + \mathrm{h.c.}\right)
       -\mu\sum_{{i} 
        \alpha}
        f^\dagger_{{i} \alpha} f_{{i} \alpha}^{}
        \\&\quad
        -B\sum_{{i} \alpha \beta}  
        \frac{1}{2} f^\dagger_{{i} \alpha} {\sigma^z_{\alpha\beta}} f_{{i} \beta}^{} ,
        \label{eq:H_MF}
    \end{aligned}
\end{equation}
where hopping $t>0$ and $\phi_{{ij}}$ is bond-dependent phase factor describing a dynamical $U(1)$ gauge field that couples to the fermions
(see Fig.~\ref{fig:trian_latt}(b)).

The total flux enclosed within a single triangular plaquette is
obtained by moving the spinon anticlockwise along three edges
of the triangle. 
In Fig.~\ref{fig:trian_latt}(b), we choose the flux configuration 
$[\theta,\pi-\theta]$ with the down-triangle flux ${\phi_\triangledown =\theta}$, 
and the up-triangle flux ${\phi_{\vartriangle}  =\pi-\theta}$.
The total number of spinons and the magnetization are controlled 
by the chemical potential $\mu$ and the Zeeman field $B$, respectively.
When ${\theta=0}$, the system is described by the U(1) DSL state 
with the $[0,\pi]$ flux configuration. 
The scalar spin-chirality term $J_\chi$ in Eq.~\eqref{eq:H_J1J2Jchi} 
induces a staggered gauge flux distribution with $\theta\neq0$ 
on top of the $[0,\pi]$ flux configuration, 
thereby driving the emergence of a CSL state 
with the $[\theta,\pi-\theta]$ flux configuration~\cite{hu2016VariationalMonte}.

\begin{figure}[t]
    \centering
    \begin{overpic}[width=0.4\linewidth]{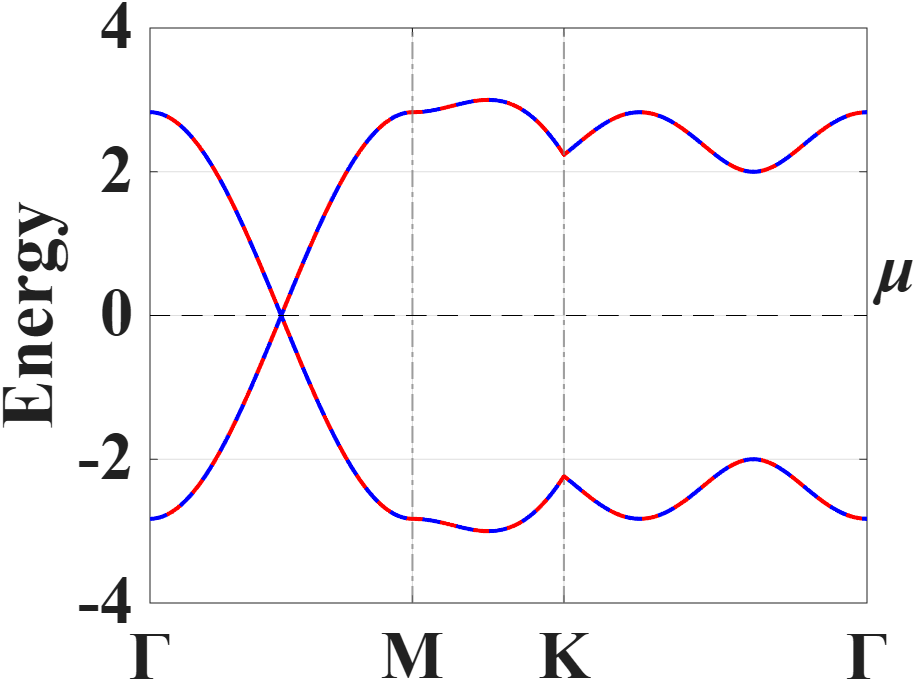}
        \put(-3,65){(a)}
    \end{overpic}
    \begin{overpic}[width=0.37\linewidth]{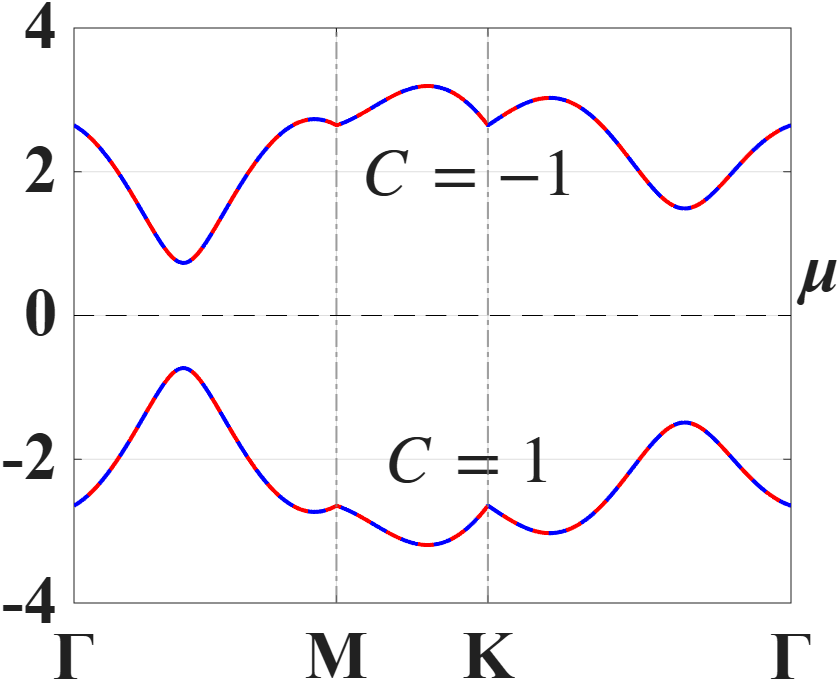}
        \put(-5,67){(b)}
    \end{overpic}
    \\
    \begin{overpic}[width=0.4\linewidth]{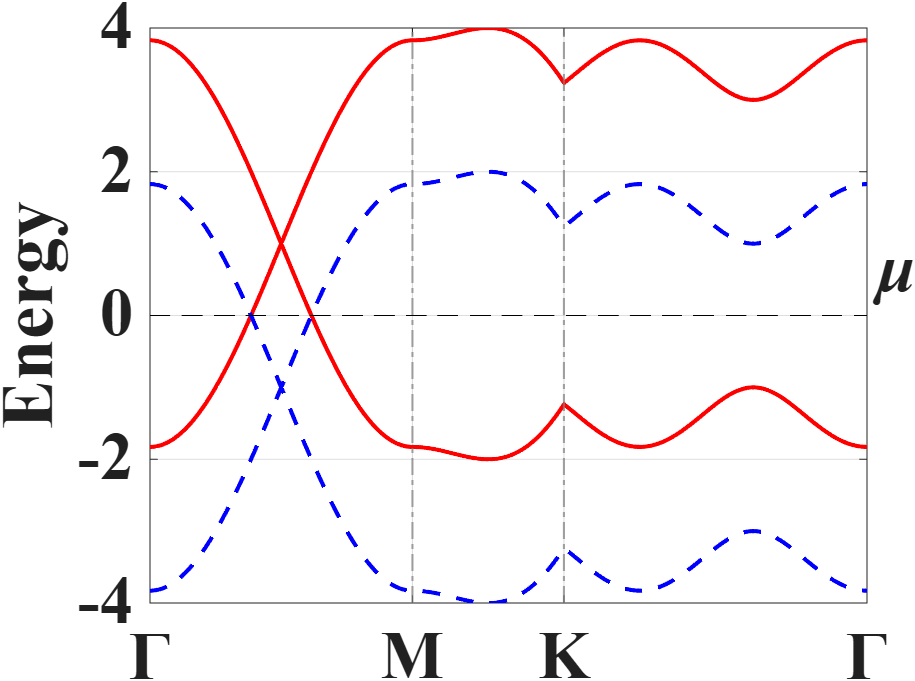}
        \put(-3,65){(c)}
    \end{overpic}
    \begin{overpic}[width=0.37\linewidth]{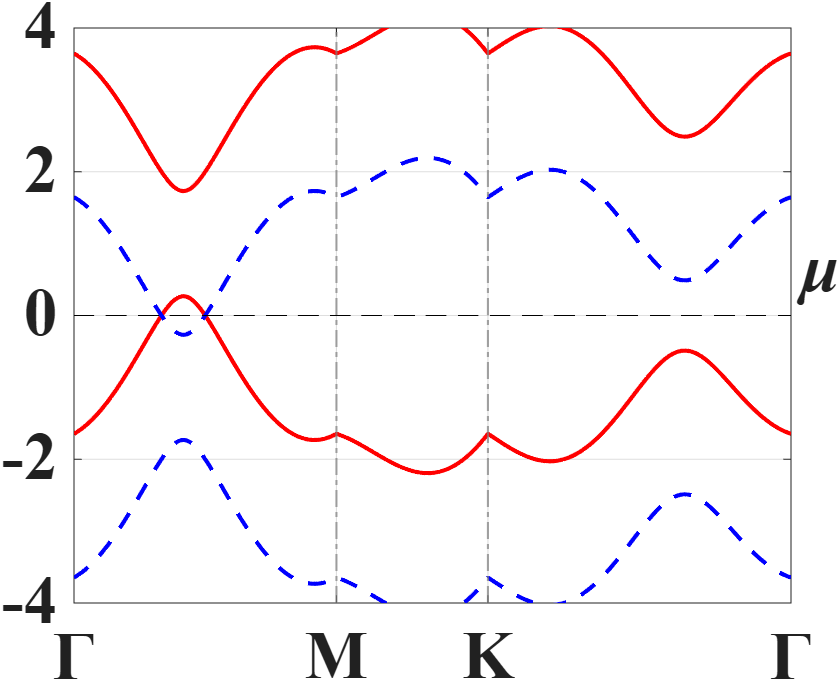}
        \put(-5,67){(d)}
    \end{overpic}
    \caption{
        Spinon dispersions of the $[\theta,\pi-\theta]$ state
        along high-symmetry lines in the Brillouin zone.
        (a) DSL with ${\theta=0}$ and (b) CSL with ${\theta=\pi/6}$ at zero field.
        The dispersions for DSL (c) and CSL (d) with ${B=2}$.
        A finite staggered flux $\theta$ gaps out the Dirac cone in (a), 
        and the resulting bands acquire nonzero Chern numbers.
        The finite $B$ splits the spin-$\uparrow$ (blue dashed line) and spin-$\downarrow$ 
        (red solid line) bands, leading to Fermi-pocket states with finite magnetizations in (c) and (d).
        }
    \label{fig:TriangleBands_B}
\end{figure}

We analyze the spinon band evolution of Eq.~\eqref{eq:H_MF} 
for the flux configuration $[\theta,\pi-\theta]$.
With ${\theta=0}$, the ground state exhibits the well-known U(1) DSL state 
with a spectrum featuring a pair of Dirac cones
for each spinon species in Fig.~\ref{fig:TriangleBands_B}(a).
Upon introducing a nonzero staggered flux $\theta$,
the system transitions into a CSL phase breaking both time-reversal and parity,
while maintaining the combined symmetry of the two.
The introduction of nonzero $\theta$ leads to
the emergence of a direct gap at each Dirac cone 
[see Fig.~\ref{fig:TriangleBands_B}(b)]. 
The lower band is fully occupied, 
while the upper band remains empty, maintaining the half-filling condition.
We find that the filled (empty) band has nonzero Chern numbers with values of $C=1(-1)$.
Taking into account the spin degeneracy, 
the net Chern number of the occupied bands amounts to $+2$, 
corresponding to a CSL phase. 
Consequently, the resulting fully gapped bands 
exhibit topologically nontrivial characteristics
such as the thermal Hall response as we will discuss in the following.

A finite Zeeman field $B$ lifts the spin degeneracy of the spinon bands 
and generates Fermi-pocket (FP) states in both the DSL and CSL phases 
in order to maintain half-filling, thereby inducing a finite magnetization, 
as shown in Fig.~\ref{fig:TriangleBands_B}(c) and (d), respectively.
For the CSL, the field-induced spin splitting needs to overcome the 
spinon gap to generate the spinon Fermi pockets. Otherwise, the CSL is robust. 
At this stage, the scalar spin chirality, that is generated by the
orbital magnetic flux, converts the DSL into the CSL, and 
the Zeeman field converts the DSL into the FP state. They 
have different effects even from this simple treatment.

\begin{figure}[t]
    \centering
    \begin{overpic}
        [width=0.4\linewidth]{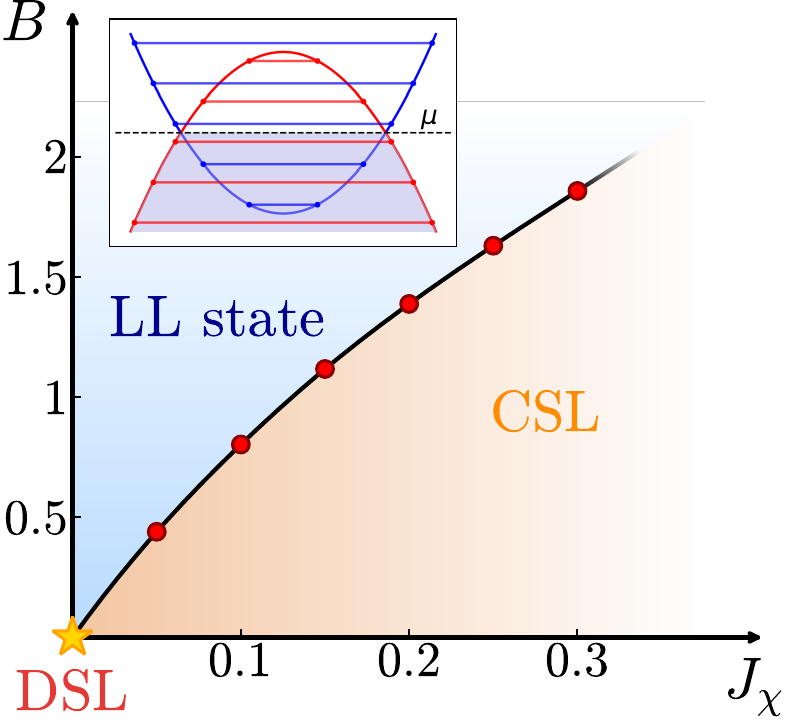}
        \put(-5,95){(a)}
        \end{overpic}
    \begin{overpic}
        [width=0.55\linewidth]{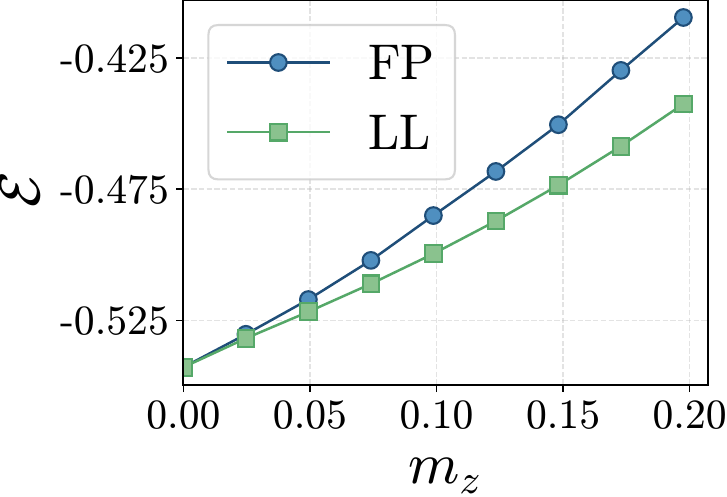}
        \put(-5,70){(b)}
    \end{overpic}%
    \caption{
        (a) Mean-field $J_\chi$-$B$ phase diagram at fixed $J_1=1$, $J_2=0.1$.
        For $J_\chi>0$, the DSL (yellow star) becomes a CSL at zero field, stable up to a critical field $B_c$ (red dots), above which a transition to the LL state occurs.
        Insets: schematic spinon band structures for the FP and LL states, with blue/red curves (levels) denoting spin-$\uparrow$/$\downarrow$ spinons.  (b) Energy density $\mathcal{E}$ as a function of magnetization density $m_z$ for FP state (blue) and LL state (green).
        }
    \label{fig:FP_LL}
\end{figure}

\noindent\bluefour{\it Spinon Landau level and spin ordering.}---Under the Zeeman coupling, 
the most direct response is spin polarization where the spinon bands are split and the system forms a FP state with finite magnetization. 
However, this FP state is not merely a passive spin-polarized state of gapless spinons. 
Because the U(1) gauge flux is dynamical, the magnetized spin liquid can further lower its energy by developing a finite uniform internal gauge flux, thereby driving the system toward a spinon Landau level (LL) state, as illustrated in Fig.~\ref{fig:FP_LL}~\cite{ran2009SpontaneousSpinOrdering,pan2025GaugeFlux,SuppMat}.
Thus the role of the Zeeman coupling is not only to polarize the spinons, but also to trigger an internal-flux instability of the emergent gauge field.
In low-energy continuum theory near the Dirac points, this instability can be demonstrated by comparing the energies of the FP state and the LL state with finite internal gauge flux, as detailed in the SM~\cite{SuppMat}. 
In the resulting magnetized spinon LL state, 
the occupied spin-$\uparrow$ Landau levels carry a total Chern number $C_\uparrow=+1$, 
while the occupied spin-$\downarrow$ Landau levels carry $C_\downarrow=-1$, yielding vanishing net Chern number and quenched thermal Hall response at zero temperature. 

The compactness of the emergent U(1) gauge field further provides a link between the finite-flux LL state and magnetic ordering. 
In a compact U(1) theory, monopole tunneling changes the total internal gauge flux by $2\pi$, and the corresponding monopole operators carry nontrivial spin and lattice quantum numbers~\cite{song2020SpinonBand,song2019UnifyingDescription,polyakov1977QuarkConfinement,polyakov2018GaugeFields}. 
For the magnetized LL state, the relevant monopole operators are $\mathcal S^{\pm}=\mathcal S_1\pm i\mathcal S_2$, formed from the elementary spin-triplet monopoles $\mathcal S_{1,2}$,
due to $\mathcal S^{\pm}\sim S^{\pm}M_{2\pi}^\dagger$ carrying $\Delta S^z=\pm 1$~\cite{song2020SpinonBand,song2019UnifyingDescription,SuppMat}.
From the lattice-symmetry analysis, the triplet monopoles transform with translation phases
$T_{1,2}:\ \mathcal S_i\to e^{\pm i2\pi/3}\mathcal S_i$,
showing that they carry momentum at the $K$ points. 
Their quantum numbers therefore identify the associated transverse magnetic ordering tendency at the $K$ points with the three-sublattice 120$^\circ$ cone pattern~\cite{song2020SpinonBand,budaraju2025MonopoleExcitations}.

\begin{figure}
    \centering
    \begin{overpic}[width=0.49\linewidth]{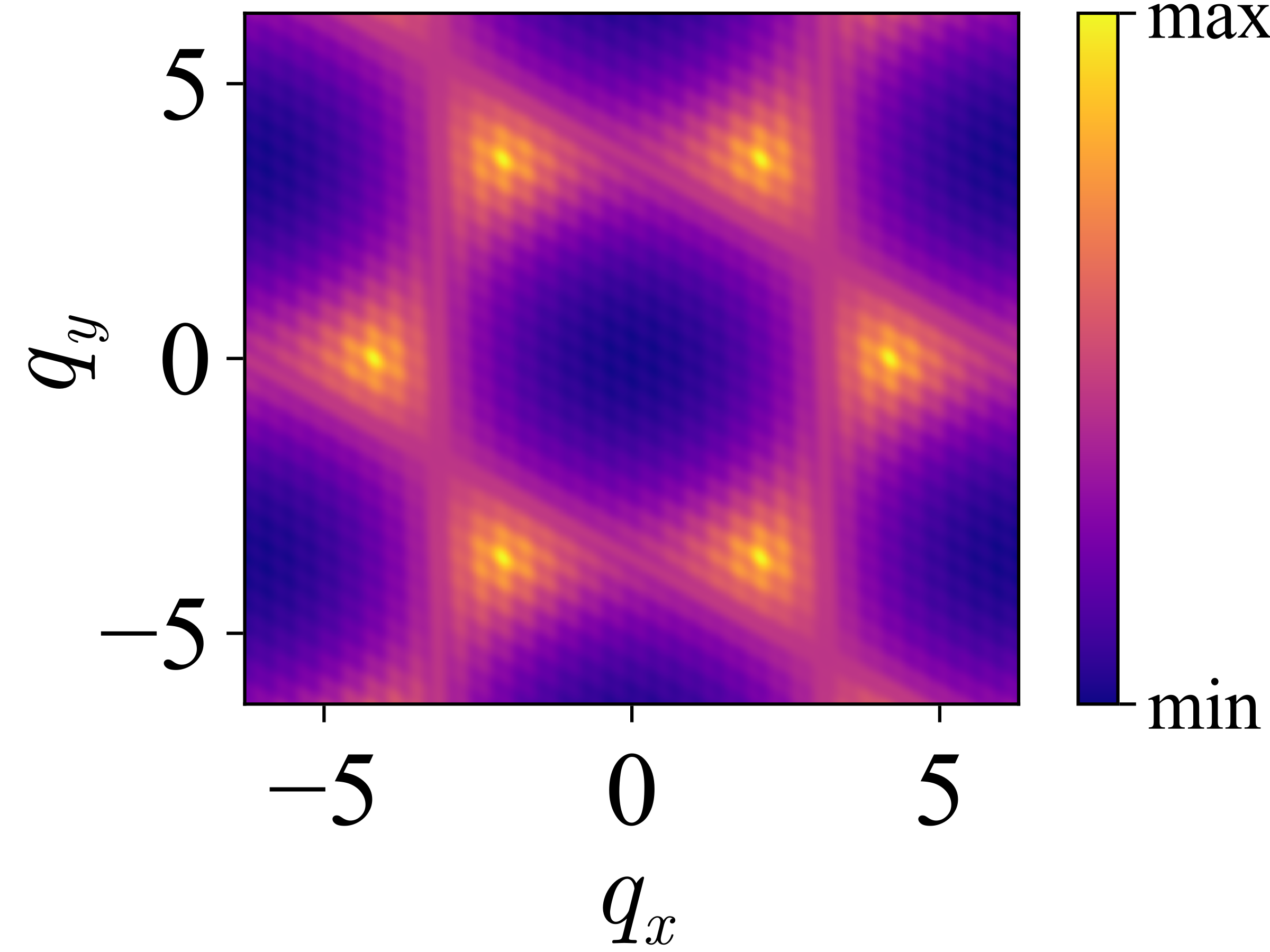}
        \put(0,75){(a)}
    \end{overpic}
    \begin{overpic}[width=0.49\linewidth]{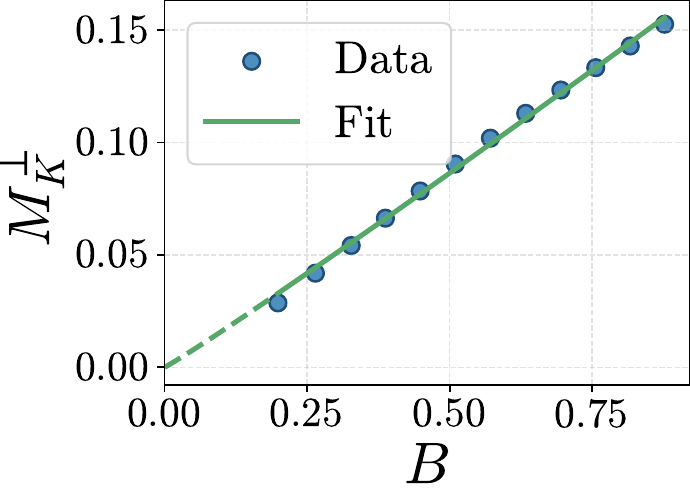}
        \put(0,75){(b)}
    \end{overpic}
    \caption{
        (a) Spin structure factor $S^{XY}(\bm{q})$ for projected LL state at ${m_z\approx0.05}$,
        where prominent peaks appear at the $K$ points of the Brillouin zone. 
        (b) The transverse magnetization $M_K^\perp$ as a function of Zeeman field $B$, 
        with scaling $M_K^\perp = 0.179 B^{1.048}$. 
     }
    \label{fig:MC_trans}
\end{figure}

To test both the energetic LL instability and the monopole-predicted transverse ordering tendency, 
we perform Gutzwiller-projected Monte Carlo calculations with gauge fluctuations included~\cite{SuppMat,becca2017QuantumMonte,gros1989PhysicsProjected}.
We construct the projected wave function 
$\ket{\Psi_n^{\text{prj}}(\{\chi_{ij}\})}=P\ket{\Psi_n^{\text{MF}}(\{\chi_{ij}\})}$, 
where $P$ is the Gutzwiller projector and 
$\ket{\Psi_n^{\text{MF}}(\{\chi_{ij}\})}$ is the mean-field wave function with $2\pi n$ flux quanta of the Hamiltonian in Eq.~\eqref{eq:H_MF}. 
We then compare different gauge-flux configurations $\{\chi_{ij}\}$ by evaluating the energy of Eq.~\eqref{eq:H_J1J2Jchi} at finite magnetization. 
On an $18 \times 18$ lattice with ${\theta=0.1\pi}$, ${J_1=1}$, ${J_2=0.1}$,
and ${J_\chi=0.3}$, the energetic preference of the LL over the FP states is shown in Fig.~\ref{fig:FP_LL}(b), providing lattice support for the uniform-flux instability found in the continuum analysis~\cite{SuppMat}.
Furthermore, we compute the transverse spin structure factor,
\begin{equation}
S^{XY}(\bm{q})=\frac{1}{N}\sum_{i,j}e^{i\bm{q}\cdot(\bm{r}_i-\bm{r}_j)}\langle S^x_i S^x_j + S^y_i S^y_j \rangle,
\end{equation}
for the projected LL state at ${m_z\approx 0.05}$ in Fig.~\ref{fig:MC_trans}(a).
The pronounced $K$ point peaks, together with finite $m_z$, signal three-sublattice 120$^\circ$ conical spin order.
This ordering is naturally associated with condensation of the spin-triplet monopole $\mathcal S^{+}$, whose quantum numbers correspond to the physical transverse-spin channel, 
giving a finite transverse moment $S^{+}\neq0$ in the spontaneously symmetry-broken ground state~\cite{song2020SpinonBand,budaraju2025MonopoleExcitations}.
Since a $2\pi$ monopole insertion connects neighboring gauge-flux sectors while changing $S^z$ by one,
the transverse magnetization $M_K^\perp$ can be extracted from the corresponding off-diagonal matrix element of the physical spin operator $S_{\mathbf r}^{+}$ between projected states in neighboring flux sectors,
\begin{equation}
    M_K^\perp=
\frac{1}{N}\sum_{\mathbf r}e^{-i\mathbf K\cdot\mathbf r}
\left\langle
\Psi_{n+1}^{\rm prj}
\left|
S_{\mathbf r}^{+}
\right|
\Psi_n^{\rm prj}
\right\rangle .
\end{equation}
For DSL, the resulting transverse moment is expected to inherit the scaling of the elementary spin-triplet monopole, giving
$M_K^\perp\sim |\langle S^+\rangle|\sim B^{\Delta_1},$
where $\Delta_1$ is the scaling dimension of the elementary spin-triplet monopole~\cite{ran2009SpontaneousSpinOrdering}.
As shown in Fig.~\ref{fig:MC_trans}(b), we obtain
$M_K^\perp=0.179B^{1.048},$
with the fitted exponent $\Delta_1=1.048$, close to the large-$N_f$ estimate $\Delta_1^{N_f}\approx 1.02$ for flavor $N_f=4$~\cite{song2019UnifyingDescription,dupuis2022AnomalousDimensions}. 
The agreement of the $K$-point spin structure and the field-scaling exponent with the monopole analysis provides further evidence that the field-induced $120^\circ$ cone order originates from condensation of the spin-triplet monopole.
These diagnostics also help distinguish the present LL scenario from a proximate valence bond solid (VBS) descendant of the triangular DSL.
The latter should primarily exhibit bond order and lattice-symmetry breaking associated with an enlarged VBS unit cell, which would manifest as additional Bragg components~\cite{song2019UnifyingDescription,song2020SpinonBand}.
Combining neutron scattering with bond-sensitive probes capable of resolving this enlarged unit cell should therefore provide a direct way to discriminate between the two scenarios.

%


\noindent\bluefour{\it Thermal Hall effect.}---The thermal Hall effect, 
capable of probing the topology of the spinon bands, 
has proven valuable for investigating the topological characteristics of the
fractionalized spinons in both experiment and theory \cite{zhao2021ThermalProperties}.
The thermal Hall conductivity for a fermionic system at finite chemical potential $\mu$ is given by~\cite{zhang2024ThermalHallEffects},
\begin{equation}
    \kappa_{xy}=-\frac{k_\mathrm{B}^2}{\hbar T}\int d\epsilon\left(\epsilon-\mu\right)^2\sigma_{xy}(\epsilon)\frac{\partial f(\epsilon,\mu,T)}{\partial\epsilon}
    \label{eq:kappa_xy}
\end{equation}
where $f(\epsilon, \mu, T )=[e^{(\epsilon-\mu)/k_BT}+1]^{-1}$ is the Fermi function,
and 
$\sigma_{xy}(\epsilon)=-\sum_{n,\epsilon_{n\boldsymbol{k}}<\epsilon}
\Omega_{n\boldsymbol{k}}(\epsilon_{n\boldsymbol{k}})$
denotes $(\hbar/e^2)$ times the zero-temperature anomalous Hall coefficient 
evaluated at chemical potential $\epsilon$. 
The Berry curvature $\Omega_{n\boldsymbol{k}}$ is defined as Kubo-like formula. 
By enforcing the single-site spinon occupancy constraint, 
the chemical potential $\mu$ is fixed such that 
the system remains at half-filling. 
The Chern number of the $n$-th band is defined as 
$C_n=\frac{1}{2\pi}\int d^2k \, \Omega_{n\boldsymbol{k}}$. 
For an isolated band, separated from all others by an energy gap, the Chern number is well defined and necessarily an integer.
In the zero-temperature limit, Eq.~\eqref{eq:kappa_xy} 
reduces to the Wiedemann--Franz form 
\begin{equation}
    \left.\frac{\kappa_{xy}}{T}\right|_{T\to0}=-\frac{\pi k_B^2}{6\hbar}\sum_{n\in\text{filled bands}}C_n.
\end{equation}
This result implies that at $T=0$, the mean-field $\kappa_{xy}/T$ is quantized in units of $\pi k_B^2/6\hbar$ for each spin species, with an additional factor of 2 in the spin-degenerate case.
After projection and gauge fluctuations are included, however, the physical low-$T$ thermal Hall response is governed by the chiral central charge $c_-$, such that $\kappa_{xy}/T=c_-\pi k_B^2/6\hbar$.
For the CSL considered here, $c_-=1$ because it is the Kalmeyer--Laughlin state, so the zero-temperature response is one quantum rather than the mean-field value of two~\cite{maity2025ThermalHall,guo2020GaugeTheories}.
Nevertheless, the temperature dependence of $\kappa_{xy}/T$ with and without gauge fluctuations should remain qualitatively similar, so the mean-field result provides a useful reference for future theoretical and experimental studies~\cite{maity2025ThermalHall}.

\begin{figure}[t]
    \centering
    \begin{overpic}
        [width=0.49\linewidth]{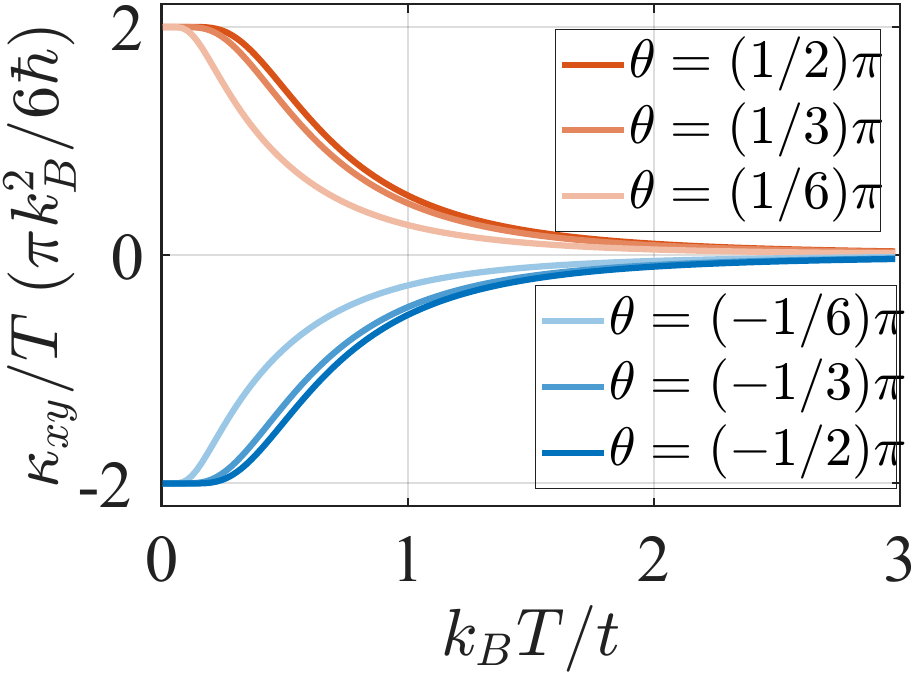}
        \put(0,77){(a)}
    \end{overpic}
    \begin{overpic}
        [width=0.49\linewidth]{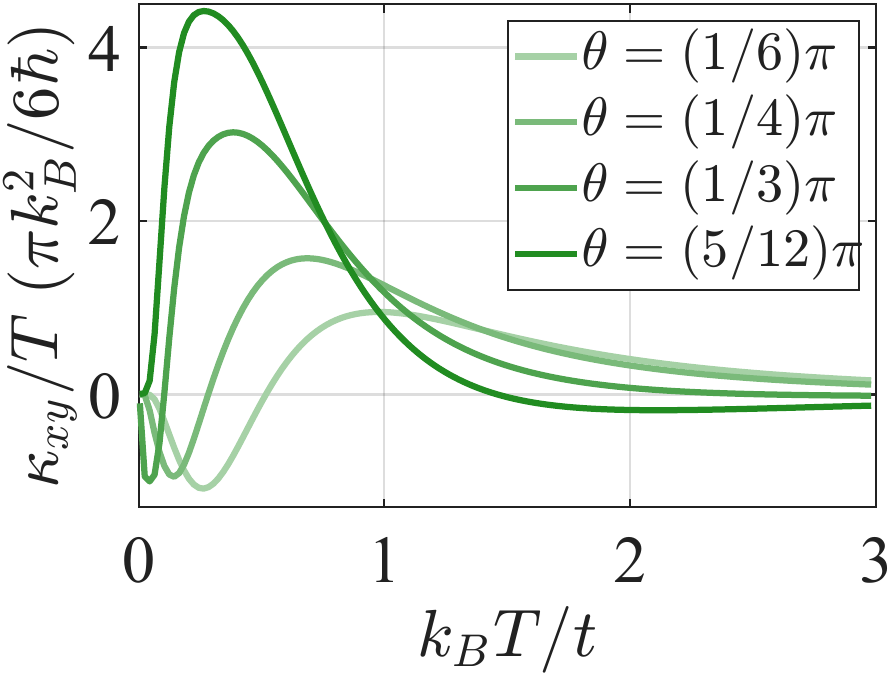}
        \put(0,77){(b)}
    \end{overpic}
    \caption{
        Temperature dependence of the spinon thermal Hall conductivity 
        $\kappa_{xy}/T$ in units of $\pi k_B^2/6\hbar$. 
        (a) Results for the CSL at ${B=0}$ and ${\phi_u=0}$ for different staggered flux $\theta$.
        As ${T \to 0}$, $\kappa_{xy}/T$ approaches the quantized value $\pm 2$, 
        and its magnitude decreases continuously toward zero with increasing temperature. 
        (b) Results for the magnetized LL state at uniform flux ${\phi_u = 2\pi/76}$ and field ${B = 1}$.
        The $\kappa_{xy}/T$ vanishes in the zero-temperature limit 
        due to the cancellation of the Chern numbers from spin-$\uparrow$ and spin-$\downarrow$ sectors.
    }
    \label{fig:kappaxy_theta}
\end{figure}

In the DSL with gapless Dirac cones, the
Brillouin-zone-integrated Berry curvature weighted by occupation 
vanishes identically due to the presence of both time-reversal and inversion symmetries, 
resulting in a zero $\kappa_{xy}/T$.
In the CSL with finite staggered flux $\theta$, both symmetries are broken, 
leading to finite spin chirality and nonzero Berry curvature distribution in the momentum space. 
Therefore the CSL phase exhibits a nontrivial thermal Hall response,
as shown in Fig.~\ref{fig:kappaxy_theta}(a).
In the CSL phase, the mean-field $\kappa_{xy}/T$ is quantized to 2 in the zero-temperature limit,
which should be reduced to 1 after including gauge fluctuations~\cite{maity2025ThermalHall}.
The mean-field quantization persists up to a finite temperature  
due to the gauge-flux-$\theta$ induced energy gap. 
It exhibits a monotonically rapid decrease due to
thermal population of spinon bands with opposite Chern number as the temperature rises. 
At temperatures significantly surpassing the maximum energy of the spinon bands, 
all the bands are almost equally populated, and the summation of Berry curvatures 
of all bands vanishes, leading to the vanishing $\kappa_{xy}/T$. 
The sign effect of the phase $\theta$ is demonstrated.
It is observed that the $\kappa_{xy}/T$ acquires a minus sign correspondingly, 
which is related to the Chern number exchanges between the occupied and unoccupied bands.

We proceed to include the spontaneously generated internal gauge flux 
from the finite Zeeman coupling $B$, and denote this uniform flux as $\phi_u$.
Consequently, the original spinon bands are split into a set of Hofstadter-like flat subbands, 
whose spin degeneracy is lifted by the Zeeman coupling \cite{SuppMat}. 
Because of the finite magnetization, the resulting spin imbalance arranges the band fillings 
such that the occupied spin-$\uparrow$ (spin-$\downarrow$) 
spinon bands carry a total Chern number of $+1$ ($-1$), yielding a vanishing net Chern number.
In Fig.~\ref{fig:kappaxy_theta}(b), we show that the thermal Hall conductivity 
$\kappa_{xy}/T$ with ${\phi_u=2\pi/76}$ and ${B=1}$ indeed approaches zero 
in the zero-temperature limit, despite the CSL parent state with a finite 
$\theta$ before a finite field is applied. 
The temperature dependence of $\kappa_{xy}/T$, however, 
is non-monotonic. For small $T$, $\kappa_{xy}/T$ first becomes negative and then rises sharply to positive values.
This behaviour reflects the presence of Hofstadter subbands with negative Chern numbers 
lying closest to the chemical potential, which becomes dominant at low temperatures.
As the temperature is further increased, the subbands with larger positive Chern numbers 
enter and contribute to the response, producing a pronounced hump in $\kappa_{xy}/T$.
At sufficiently high temperatures, all bands are thermally populated and their 
Berry-curvature contributions cancel, causing $\kappa_{xy}/T$ to decay toward zero. 

As $\theta$ approaches zero, the thermal Hall response diminishes gradually and 
vanishes entirely across all temperatures when ${\theta=0}$.
This occurs because contributions from spin-$\uparrow$ and spin-$\downarrow$ spinons can cancel each other at all temperatures in the LL state derived from the DSL at ${\theta=0}$~\cite{SuppMat}.
Specifically, the total Hall coefficient $\sigma_{xy}^{\text{tot}}$ is an odd function of energy due to the symmetric properties of the Berry curvatures.
Upon integrating over energy with the remaining even-function factors in Eq.~\eqref{eq:kappa_xy}, the thermal Hall conductivity $\kappa_{xy}/T$ vanishes identically at all temperatures.
The distinct thermal Hall conductivity behaviors exhibited by the DSL, CSL, and LL states
under combined effect of Zeeman field $B$ and internal gauge flux $\phi_u$ thus provide useful experimental signatures 
to distinguish these exotic quantum states.


\noindent\bluefour{\it Discussion.}---We note that since $C_\uparrow+C_\downarrow=0$, no Chern-Simons term is generated for the emergent gauge field, 
consistent with the vanishing thermal Hall conductance of the LL state at zero temperature. 
The gauge field therefore retains a gapless Maxwell-like mode, which in the cone-ordered phase can be identified with the Goldstone mode of the spontaneously broken spin $U(1)$ symmetry~\cite{ran2009SpontaneousSpinOrdering}, 
which should appear in low-energy spin correlations accessible to neutron scattering. 
If a conventional bosonic thermal Hall response occurs in the field-induced cone state, it should instead be a nonuniversal finite-temperature effect arising from Berry curvature of thermally populated magnon bands. 
From this perspective, Schwinger-boson approaches provide a complementary description of the magnetic order, its collective excitations, and the possible magnonic thermal Hall response~\cite{zhang2024ThermalHallEffects}.

In moir\'e systems, the valley or layer degree of freedom can play the role of spin, and once this pseudospin forms an in-plane ordered state, it can give rise to unique electronic polarization which should be detectable~\cite{kuhlenkamp2024ChiralPseudospin}.
And our conical state with scaling $M_K^\perp\sim B^{\Delta_1}$ is different from the conventional coplanar antiferromagnetic order, where the in-plane order remains nonzero for $B=0$ and can be canted or rotated for $B>0$.

For Yb-based and Co-based triangular antiferromagnets,
the DSL physics has also been proposed to be relevant~\cite{jia2024QuantumSpin,lee2021TemporalField,zhong2019StrongQuantum,
zhong2020FrustratedMagnetism,zhu2024ContinuumExcitations,shen2016EvidenceSpinon,
shen2018FractionalizedExcitations,bag2024EvidenceDiracQuantum,wu2025SpinDynamics}.
As these materials lie in the strong Mott regime, the spin-liquid physics arises 
primarily from competing pairwise interactions, and the flux term can be neglected. 
Moreover, the finite-temperature thermally disordered regime of the 
weakly ordered states, which lies in proximity to the nearby spin liquid~\cite{jia2024QuantumSpin},
may be more appropriately described in terms of a thermalized spin liquid. 
The magnetic-field response and related properties in this regime may then be understood 
from thermalized LL states. It would be interesting to explore this direction 
both numerically and experimentally.

To summarize, we investigate the triangular spin liquids using a parton mean-field approach 
that incorporates both the Zeeman effect and the internal U(1) gauge flux. We show that orbital flux favors the CSL, whereas Zeeman coupling destabilizes the spinon FP state toward a LL state with spontaneous uniform internal flux and conical spin order. These states exhibit distinct thermal Hall and spin-structure-factor signatures, providing concrete probes for moir\'e Hofstadter-Hubbard systems and triangular-lattice antiferromagnets.

\noindent\bluefour{\it Note added.}---Recent independent works have reported related field-induced monopole physics in the triangular-lattice $J_1$-$J_2$ model, including finite-flux states with dominant 120$^\circ$ correlations~\cite{wang2026ChiralityQuasilongrange} and a monopole phase with finite scalar chirality~\cite{budaraju2026EmergenceMonopole}.

\noindent\bluefour{\it Acknowledgments.}---J.Y. thanks Wei Zhu, Xu-Ping Yao, Zheng-Xin Liu, Xuan-Zhe Xia 
for helpful discussions.
This work is supported by NSFC with Grants No.92565110 and No.12574061,
by the Ministry of Science and Technology of China with Grants No. 2021YFA1400300, 
and by the Fundamental Research Funds for the Central Universities, Peking University.

\bibstyle{apsrev-nourl}
\bibliography{Refs}



\clearpage
\newpage
\onecolumngrid

\begin{center}
\textbf{\large Supplemental Material for \\
``Emergent Gauge Flux and Spin Ordering in Magnetized Triangular Spin Liquids:\\
Applications to Hofstadter-Hubbard Model''
}
\end{center}

\addtocontents{toc}{\protect\setcounter{tocdepth}{0}}
{
\tableofcontents
}

\renewcommand{\thefigure}{S\arabic{figure}}
\setcounter{figure}{0}
\renewcommand{\theequation}{S\arabic{equation}}
\setcounter{equation}{0}
\renewcommand{\thesection}{\Roman{section}}
\setcounter{section}{0}
\setcounter{secnumdepth}{4}


\section{Effective spin model}
\label{app:effective spin model}

We derive the effective spin Hamiltonian up to order $t^4/U^3$ for the Hofstadter-Hubbard model with complex hopping amplitudes $t_{ij} = t e^{i\varphi_{ij}}$, where the phase $\varphi_{ij}$ encodes the Peierls phase induced by the orbital magnetic flux.
Considering the Hamiltonian of the Hofstadter-Hubbard model,
\begin{equation}
    \mathcal{H}=-\sum_{\langle ij\rangle,\sigma}t_{ij}c_{i\sigma}^\dagger c_{j\sigma}+U\sum_in_{i\uparrow}n_{i\downarrow},
\end{equation}
where $\langle ij\rangle$ denotes nearest-neighbor hopping, $\sigma$ is the spin degree of freedom, and $U$ is the strength of the local Hubbard repulsion.
In the strong-coupling limit ($U\gg t$),
the effective Hamiltonian to order $t^4 / U^3$ on the isotropic triangular lattice reads \cite{motrunich2006OrbitalMagneticField,delannoy2005NeelOrder,cookmeyer2021FourSpinTerms}
\begin{equation}
\begin{aligned}
\mathcal{H}_{\mathrm{eff}}&=\mathcal{H}_{\mathrm{eff}}^{(2)}+\mathcal{H}_{\mathrm{eff}}^{(3)}+\mathcal{H}_{\mathrm{eff}}^{(4)},
\\\mbox{with}\quad
\mathcal{H}_{\mathrm{eff}}^{(2)}&=J_1^{(2)}\sum_{\langle ij\rangle}\bm{S}_i\cdot\bm{S}_j,
\\\mathcal{H}_{\mathrm{eff}}^{(3)}&=J_{\chi}
\sum_{\vartriangle}
\bm{S}_1\cdot\bm{S}_2\times\bm{S}_3,
\\\mathcal{H}_{\mathrm{eff}}^{(4)}&=J_1^{(4)}\sum_{\langle ij\rangle}\bm{S}_i\cdot\bm{S}_j+J_2^{(4)}\sum_{\langle\langle ij\rangle\rangle}\bm{S}_i\cdot\bm{S}_j
\\&
+J_3^{(4)}\sum_{\langle\langle\langle ij\rangle\rangle\rangle}\bm{S}_i\cdot\bm{S}_j+J_{\bm{R}}^{(4)}\sum_{\langle i,j,k,l\rangle}\mathcal{R}_{ijkl},
\\\mbox{ and}\quad
\mathcal{R}_{ijkl}&=(\bm{S}_i\cdot\bm{S}_j)(\bm{S}_k\cdot\bm{S}_l)+(\bm{S}_i\cdot\bm{S}_l)(\bm{S}_j\cdot\bm{S}_k)
-(\bm{S}_i\cdot\bm{S}_k)(\bm{S}_j\cdot\bm{S}_l).
\end{aligned}
\end{equation}
The antiferromagnetic interactions
$J_1^{(2),(4)}$, $J_2^{(4)}$, $J_3^{(4)}$, $J_\chi$ and $J_\mathrm{R}^{(4)}$ denote the first-, second-, third-neighbor and 3-site-ring, 4-site-ring exchange couplings, with
\begin{equation}
\begin{aligned}
J_1^{(2)}&=\frac{4t^2}{U}, 
\quad
J_1^{(4)}=-\frac{24t^4}{U^3},
\quad 
J_2^{(4)}=J_3^{(4)}=\frac{4t^2}{U^3},
\\J_{\chi}&=-\frac{24t^3}{U^2}\sin(\Phi_{\vartriangle}),
\quad
J_{R}^{(4)}=\frac{80t^4}{U^3}\cos(\Phi_{\lozenge}).
\end{aligned}
\end{equation}
where dimensionless flux $\Phi_{\vartriangle}$ is defined as the sum of the Peierls phases around the triangle $\vartriangle$ and {$\Phi_{\vartriangle}=\Phi_{\lozenge}/2$ }.

In reality, one can apply the external magnetic field 
away from the $z$ direction. The $z$ component component 
contributes to the orbital magnetic flux, and the total magnetic field 
contributes to the Zeeman coupling. 
As the effective spin model with the ring exchange 
has the global SU(2) symmetry, the presence of the Zeeman coupling
in the effective model breaks the SU(2) symmetry down to U(1),
and then the Zeeman coupling can be conveniently set to be along $z$ direction
in the spin space.

\subsection*{Mean-field phase diagram}

We perform the mean-field calculations of the effective spin model $\mathcal{H}_{\mathrm{eff}}$ with $J_1$, $J_2$ and $J\chi$, 
using the fermionic parton construction described in the main text.
The mean-field phase diagram in the $J_\chi$-$B$ plane with fixed $J_1=1$ and $J_2=0.1$ is shown in Fig.~3(b).
When both $J_\chi$ and $B$ vanish, the ground state corresponds to a U(1) DSL~\cite{wietek2017ChiralSpinLiquid,gong2017GlobalPhaseDiagram}.
In the absence of magnetic field, increasing $J_\chi$ from zero drives the DSL into a CSL.
This CSL remains stable up to a critical field $B_c$, above which the system transitions to the LL state.
The staggered flux parameter $\theta$ for different $J_\chi$ at phase boundary is in agreement with the variational Monte Carlo results of Ref.~\cite{hu2016VariationalMonte}, 
and the critical field $B_c$ is determined by the closing of the spinon gap in the CSL phase.

\section{Elementary monopoles in $N_f=4$ QED$_3$}

The low-energy fixed-point theory of the DSL is described by $N_f=4$ QED$_3$. When monopole tunneling events are neglected, the emergent $U(1)$ gauge field is treated as noncompact and gauge-flux conservation gives rise to an emergent $U(1)_{\mathrm{top}}$ symmetry. Microscopic compactness is encoded by local monopole operators that insert $\pm 2\pi$ gauge flux. Their spin and lattice symmetry quantum numbers identify the physical ordering channels associated with monopole proliferation~\cite{song2019UnifyingDescription,song2020SpinonBand}.

To construct the elementary monopoles, it is useful to think in the large-$N_f$ limit or, equivalently, through the state-operator correspondence on $S^2$. In a $2\pi$ flux background, each Dirac flavor contributes one fermion zero mode, so for $N_f=4$ there are four zero modes total. Gauge neutrality requires filling exactly half of them, hence there are
\begin{equation}
\binom{4}{2}=6
\end{equation}
elementary monopoles. Because the zero modes in a $2\pi$ flux background carry no additional Lorentz spin, these elementary monopoles are Lorentz scalars. They organize into three spin singlets $\mathcal V_{1,2,3}$ and three spin triplets $\mathcal S_{1,2,3}$.
In the zero-mode basis $f^\dagger_{1\uparrow},f^\dagger_{1\downarrow},f^\dagger_{2\uparrow},f^\dagger_{2\downarrow}$ one may write, up to overall normalization and convention-dependent phases,
\begin{align}
\mathcal V_1^\dagger &\sim \left(f^\dagger_{1\uparrow}f^\dagger_{1\downarrow}-f^\dagger_{2\uparrow}f^\dagger_{2\downarrow}\right)\mathcal M_{bare}^\dagger, \nonumber\\
\mathcal V_2^\dagger &\sim i\left(f^\dagger_{1\uparrow}f^\dagger_{1\downarrow}+f^\dagger_{2\uparrow}f^\dagger_{2\downarrow}\right)\mathcal M_{bare}^\dagger, \nonumber\\
\mathcal V_3^\dagger &\sim \left(f^\dagger_{1\downarrow}f^\dagger_{2\uparrow}-f^\dagger_{1\uparrow}f^\dagger_{2\downarrow}\right)\mathcal M_{bare}^\dagger,
\end{align}
and
\begin{align}
\mathcal S_1^\dagger &\sim i\left(f^\dagger_{1\uparrow}f^\dagger_{2\uparrow}-f^\dagger_{1\downarrow}f^\dagger_{2\downarrow}\right)\mathcal M_{bare}^\dagger, \nonumber\\
\mathcal S_2^\dagger &\sim -\left(f^\dagger_{1\uparrow}f^\dagger_{2\uparrow}+f^\dagger_{1\downarrow}f^\dagger_{2\downarrow}\right)\mathcal M_{bare}^\dagger, \nonumber\\
\mathcal S_3^\dagger &\sim -i\left(f^\dagger_{1\uparrow}f^\dagger_{2\downarrow}+f^\dagger_{1\downarrow}f^\dagger_{2\uparrow}\right)\mathcal M_{bare}^\dagger.
\end{align}
The six elementary monopoles form the vector representation of the emergent flavor symmetry $SO(6)$ and carry unit $U(1)_{\mathrm{top}}$ charge, so the proper continuum symmetry structure is $(SO(6)\times U(1)_{\mathrm{top}})/\mathbb{Z}_2$~\cite{song2020SpinonBand}. A useful point emphasized by Song \emph{et al.} is that monopoles are local gauge-invariant operators, so they transform under ordinary linear representations of the microscopic symmetry group, in contrast to the spinons which transform projectively.

The group-theory content is then transparent. The operators $\mathcal S_{1,2,3}$ transform as the Cartesian components of a physical-spin triplet, while $\mathcal V_{1,2,3}$ are spin singlets and form a triplet in the valley sector. In a spherical basis,
\begin{align}
\mathcal S_{+1}^\dagger &\sim f^\dagger_{1\uparrow}f^\dagger_{2\uparrow}\mathcal M_{bare}^\dagger, \nonumber\\
\mathcal S_0^\dagger &\sim \frac{f^\dagger_{1\uparrow}f^\dagger_{2\downarrow}+f^\dagger_{1\downarrow}f^\dagger_{2\uparrow}}{\sqrt{2}}\mathcal M_{bare}^\dagger, \nonumber\\
\mathcal S_{-1}^\dagger &\sim f^\dagger_{1\downarrow}f^\dagger_{2\downarrow}\mathcal M_{bare}^\dagger.
\end{align}
By contrast, $\mathcal V_{1,2,3}$ are physical-spin singlets, but they have the same algebraic structure as a triplet in the valley pseudospin sector:
\begin{align}
\mathcal V_{+1}^\dagger &\sim f^\dagger_{1\uparrow}f^\dagger_{1\downarrow}\mathcal M_{bare}^\dagger, \nonumber\\
\mathcal V_0^\dagger &\sim \frac{f^\dagger_{1\uparrow}f^\dagger_{2\downarrow}-f^\dagger_{1\downarrow}f^\dagger_{2\uparrow}}{\sqrt{2}}\mathcal M_{bare}^\dagger, \nonumber\\
\mathcal V_{-1}^\dagger &\sim f^\dagger_{2\uparrow}f^\dagger_{2\downarrow}\mathcal M_{bare}^\dagger.
\end{align}
Thus, the spin-triplet monopoles furnish the transverse magnetic-ordering channels. With the conventions above,
\begin{equation}
\mathcal S_{+1}^{\dagger}\propto
\mathcal S_1^{\dagger}-i\mathcal S_2^{\dagger},
\qquad
\mathcal S_{-1}^{\dagger}\propto
\mathcal S_1^{\dagger}+i\mathcal S_2^{\dagger},
\end{equation}
which carry $\Delta S^z=+1$ and $-1$, respectively. A Zeeman term $H_Z=-h\sum_{\mathbf r}S_{\mathbf r}^{z}$ preserves the residual $U(1)_{S^z}$ symmetry and therefore cannot linearly source either of these transverse monopole operators. Instead, it produces a spin imbalance and changes the relative energetics of the spinful flux sectors, thereby potentially favoring one transverse ordering channel over the other.

For the triangular lattice, however, the zero-mode construction alone does not fully determine the monopole quantum numbers. A microscopic symmetry acts on a monopole in two pieces: an $SO(6)$ rotation inherited from the fermion projective symmetry group, and an additional $U(1)_{\mathrm{top}}$ phase coming from the Berry phase of the filled Dirac sea~\cite{song2019UnifyingDescription,song2020SpinonBand}. The latter is the subtle part. 
Ref.~\cite{song2020SpinonBand} fixes it analytically by deforming the DSL to a gapped quantum-spin-Hall state and determining the charge-center, or Wannier-center, structure of the occupied spinon bands. For the triangular staggered-$\pi$-flux ansatz, the occupied band can be represented schematically as
$\textit{occupied band}=\textit{site}+\vartriangle-\triangledown$,
which, after combining with the background gauge charge $-1$ per site imposed by the single-occupancy constraint, implies that the spin-triplet monopoles see zero net gauge charge for site-centered $C_3$ rotations but opposite effective charges at the up- and down-triangle centers. As a result, the triplet monopoles acquire opposite $\pm 2\pi/3$ Berry phases under the two triangle-centered $C_3$ rotations. Using the algebraic relations between rotations and translations, one then obtains the triangular-lattice translation quantum numbers
\begin{equation}
T_1:\ \mathcal{S}^{\dagger}_{1,2,3}\to e^{i2\pi/3}\mathcal{S}^{\dagger}_{1,2,3},
\qquad
T_2:\ \mathcal{S}^{\dagger}_{1,2,3}\to e^{-i2\pi/3}\mathcal{S}^{\dagger}_{1,2,3},
\end{equation}
showing that the elementary spin-triplet monopoles carry crystal momentum $K$ rather than $\Gamma$~\cite{song2020SpinonBand}. 
Ref.~\cite{song2019UnifyingDescription} further confirmed this assignment numerically on a torus by uniformly spreading the $2\pi$ flux and directly extracting the monopole momentum from symmetry overlaps.
Since time reversal reverses the emergent gauge flux, it maps a monopole to the corresponding antimonopole. In the same convention,
\begin{equation}
\mathcal T:\quad
\mathcal S_i^\dagger\to-\mathcal S_i,
\qquad
\mathcal V_i^\dagger\to\mathcal V_i,
\qquad i=1,2,3.
\end{equation}
Thus, the triplet and singlet monopoles differ by their relative time-reversal sign~\cite{song2019UnifyingDescription,song2020SpinonBand}.

The elementary $2\pi$ spin-triplet monopoles transform as spin-1 operators carrying crystal momentum $K$. Consequently, a condensate in the $\mathcal S_{\pm1}^{\dagger}$ monopole channel transforms in the same microscopic symmetry representation as a transverse $K$-point magnetic order parameter. It is therefore associated, by symmetry, with the three-sublattice $120^\circ$ ordering channel.

To make this correspondence explicit, define the complex monopole condensate
\begin{equation}
\boldsymbol{\Phi}_{M}
\equiv
\left\langle
\left(
\mathcal S_1^\dagger,
\mathcal S_2^\dagger,
\mathcal S_3^\dagger
\right)
\right\rangle
=\mathbf n_1-i\mathbf n_2,
\end{equation}
where $\mathbf n_1=\operatorname{Re}\boldsymbol{\Phi}_{M}$ and $\mathbf n_2=-\operatorname{Im}\boldsymbol{\Phi}_{M}$. Because $\boldsymbol{\Phi}_{M}$ and the physical transverse magnetic order parameter transform identically under spin and lattice symmetries, the corresponding real-space spin texture may be parameterized as
\begin{equation}
\left\langle\mathbf S_{\mathbf r}\right\rangle_{\perp}
=
\lambda\,\operatorname{Re}
\left[
\boldsymbol{\Phi}_{M}e^{i\mathbf K\cdot\mathbf r}
\right]
=
\lambda
\left[
\mathbf n_1\cos(\mathbf K\cdot\mathbf r)
+\mathbf n_2\sin(\mathbf K\cdot\mathbf r)
\right],
\end{equation}
where $\lambda$ is a nonuniversal proportionality coefficient.

For a condensate purely in the $\Delta S^z=+1$ channel, $\langle\mathcal S_{+1}^{\dagger}\rangle\neq0$ while $\langle\mathcal S_{-1}^{\dagger}\rangle=\langle\mathcal S_{0}^{\dagger}\rangle=0$. This implies $\boldsymbol{\Phi}_{M}\cdot\boldsymbol{\Phi}_{M}=0$, or equivalently $\mathbf n_1\cdot\mathbf n_2=0$ and $|\mathbf n_1|=|\mathbf n_2|$. The transverse spin amplitude is therefore constant, giving a circular rather than elliptical spiral. Since the phase advances by $2\pi/3$ among the three sublattices at $\mathbf K$, this is the coplanar $120^\circ$ state. When the Zeeman field additionally induces a uniform magnetization $m_z$, the full texture becomes
\begin{equation}
\left\langle\mathbf S_{\mathbf r}\right\rangle
=
\left\langle\mathbf S_{\mathbf r}\right\rangle_{\perp}
+m_z\hat{\mathbf z},
\end{equation}
which is the conical, or umbrella, state.

\section{Continuum theory of spinon Landau level instability}

Based on aforementioned discussions, the low-energy physics can be effectively described by Dirac cones near the chemical potential.
For illustration, we take the low-energy continuum theory of Dirac fermion with gauge field~\cite{song2021DopingChiral}.
We compare the Fermi-pocket (FP) and Landau-level (LL) states at fixed spin
imbalance density $\Delta n=(N_{\uparrow}-N_{\downarrow})/A$,
which is the same quantity denoted by the magnetization density $m_z$ in the
main text. At fixed $\Delta n$, the Zeeman contribution is common to the two
trial states and will be omitted. The following continuum formulas keep the two
Dirac nodes of the spinon spectrum explicitly.


For each Dirac node, the massless spinon dispersion is $\epsilon_{\pm}(\bm{k})=\pm v_F|\bm{k}|$.
The Zeeman field creates one spin-$\uparrow$ particle pocket and one
spin-$\downarrow$ hole pocket at each node. If their common radius is $k_F$,
then the state counting gives
\begin{equation}
\Delta n
=
2_{\rm nodes}\times 2_{\rm particle/hole}
\frac{k_F^2}{4\pi}
=
\frac{k_F^2}{\pi}.
\end{equation}
The particle and hole pockets have the same kinetic energy cost relative to the
filled Dirac sea. Summing the four pockets and using $k_F^2=\pi\Delta n$ gives
\begin{equation}
\Delta e_{MF}^{FP}
=
4\int_{|\bm{k}|<k_F}\frac{d^2 k}{(2\pi)^2}\,v_F|\bm{k}|
=
\frac{2v_Fk_F^3}{3\pi}
=
\frac{2\sqrt{\pi}v_F}{3}(\Delta n)^{3/2}.
\end{equation}


Now allow a uniform internal gauge flux $b$. For one node and one spin species,
the relativistic Landau levels are
\begin{equation}
E_{n,\pm}=\pm v_F\sqrt{2nb},
\qquad n=1,2,\ldots,
\end{equation}
with a zero-energy Landau level $E_0=0$ and degeneracy $b/(2\pi)$ per unit area.
The LL state carries the spin imbalance by filling the spin-$\uparrow$ zeroth
Landau levels of the two Dirac nodes, so
\begin{equation}
\Delta n=2_{\rm nodes}\frac{b}{2\pi}=\frac{b}{\pi}.
\end{equation}
The zeroth level itself costs no kinetic energy in the massless theory. The
remaining contribution is the zeta-regularized change of the filled Dirac sea,
which for the four Dirac flavors gives~\cite{ran2009SpontaneousSpinOrdering}
\begin{equation}
\Delta e_{MF}^{LL}
=
4v_F\left[
-\frac{b}{2\pi}\sum_{n=1}^{\infty}\sqrt{2nb}
-\int\frac{d^2 k}{(2\pi)^2}(-|\bm{k}|)
\right]
=
\frac{\zeta\left(\frac{3}{2}\right)}{\sqrt{2}\pi^2}
 v_F b^{3/2}
=
\frac{\zeta\left(\frac{3}{2}\right)}{\sqrt{2\pi}}
 v_F(\Delta n)^{3/2}.
\end{equation}


For a massive Dirac spectrum
\begin{equation}
E_{\pm}(\bm{k})=\pm\sqrt{M^2+v_F^2|\bm{k}|^2},
\qquad M>0,
\end{equation}
the pocket counting is unchanged: $\Delta n=k_F^2/\pi$. The massive FP energy
therefore becomes
\begin{equation}
\Delta e_{MF}^{MFP}
=
4\int_{|\bm{k}|<k_F}
\frac{d^2k}{(2\pi)^2}
\sqrt{M^2+v_F^2|\bm{k}|^2}
=
\frac{2}{3\pi v_F^2}
\left[
\left(M^2+\pi v_F^2\Delta n\right)^{3/2}-M^3
\right].
\end{equation}
It is useful to introduce
\begin{equation}
\lambda_{\Delta}=\frac{M^2}{2\pi v_F^2\Delta n},
\qquad
\mathcal G(\lambda)
=
\frac{2\sqrt{\pi}}{3}
\left[
\left(1+2\lambda\right)^{3/2}
-
\left(2\lambda\right)^{3/2}
\right],
\end{equation}
so that
\begin{equation}
\Delta e_{MF}^{MFP}=v_F(\Delta n)^{3/2}\mathcal G(\lambda_{\Delta}).
\end{equation}


In a uniform internal flux $b>0$, the paired massive Landau levels are
\begin{equation}
E_{n,\pm}=\pm\sqrt{M^2+2nv_F^2b},
\qquad n=1,2,\ldots,
\end{equation}
while the zeroth Landau levels sit at energies $E_{0,a}=-M$ for the two nodes. The degeneracy is still $b/(2\pi)$ per node and spin, so the
LL density relation remains $\Delta n=b/\pi$.

The paired-level Dirac-sea contribution can be written compactly as
\begin{equation}
\Delta e_{\ge1}^{MLL}
=4
\left[
-\frac{b}{2\pi}
\sum_{n=1}^{\infty}
\sqrt{M^2+2nv_F^2b}
+\int\frac{d^2k}{(2\pi)^2}
\sqrt{M^2+v_F^2|\bm{k}|^2}
\right]
=
2\sqrt{2\pi}\,v_F(\Delta n)^{3/2}
\mathcal F(\lambda_{\Delta}),
\end{equation}
where $\mathcal F(\lambda)
=-\frac{2}{3}\lambda^{3/2}-
\zeta\left(-\frac{1}{2},1+\lambda\right)$.
The occupied zeroth Landau levels contribute
\begin{equation}
\Delta e_0^{MLL}=
\frac{b}{2\pi}\sum_{a=1}^{2}E_{0,a}
=-2\times
\frac{M\Delta n}{2}=- M\Delta n.
\end{equation}
Thus
\begin{equation}
\Delta e_{MF}^{MLL}
=
v_F(\Delta n)^{3/2}
\left[
2\sqrt{2\pi}\,\mathcal F(\lambda_{\Delta})
-\sqrt{2\pi\lambda_{\Delta}}
\right].
\end{equation}


For the massless comparison,
\begin{equation}
\Delta e_{MF}^{FP}
=
\frac{2\sqrt{\pi}v_F}{3}(\Delta n)^{3/2},
\qquad
\Delta e_{MF}^{LL}
=
\frac{\zeta\left(\frac{3}{2}\right)}{\sqrt{2\pi}}
 v_F(\Delta n)^{3/2}.
\end{equation}
Since
\begin{equation}
\frac{\Delta e_{MF}^{LL}}{\Delta e_{MF}^{FP}}
=
\frac{3\zeta\left(\frac{3}{2}\right)}{2\sqrt{2}\pi}<1,
\label{eq:ratio}
\end{equation}
the massless LL state is lower than the FP state at mean-field level.

For nonzero $M$, the LL state is lower than the massive FP state, since we can numerically check this relation holds for all $\lambda_{\Delta}>0$:
\begin{equation}
\frac{\Delta e_{MF}^{MLL}}{\Delta e_{MF}^{MFP}}
=\frac{2\sqrt{2\pi}\,\mathcal F(\lambda_{\Delta})
-\sqrt{2\pi\lambda_{\Delta}}}{\mathcal G(\lambda_{\Delta})}
<
1.
\end{equation}

\section{Uniform flux construction in lattice Hamiltonian}
\label{sm:Latt.UFlux}

\subsection{Landau gauge}
\label{sm:LandauGauge}

To incorporate a uniform flux $\phi_u$ per triangle into the triangular lattice, we can choose the Landau gauge. In this gauge, the nontrivial Peierls phases for nearest-neighbor hopping are defined as $\varphi_1(i)=\phi_u(2\cdot \bm{r}_i-1/2)$ along the $\bm{a}_2$ direction, and $\varphi_2(i)=\phi_u(2\cdot \bm{r}_i+1/2)$ along the $\bm{a}_3=\bm{a}_2-\bm{a}_1$ direction \cite{du2018FloquetHofstadterButterfly}.
With coprime integers $p$ and $q$, the uniform flux $\phi_u = 2\pi\,p/(2q)$ threads each elementary triangle (either $\vartriangle$ or $\triangledown$), resulting in an enlarged magnetic unit cell along the $\bm{a}_1$ direction that contains $q$ sites.

\begin{figure}[t]
    \centering
    \begin{overpic}
        [width=0.35\linewidth]{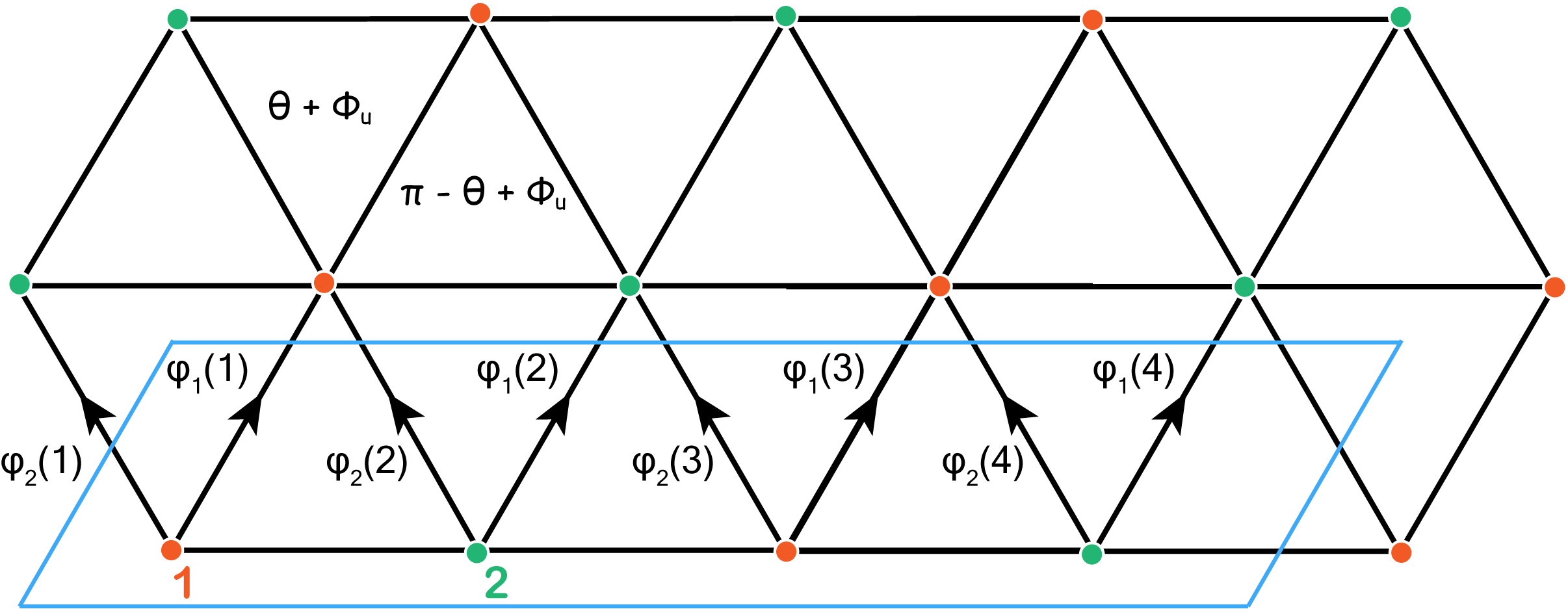}
        \put(0,35){(a)}
    \end{overpic}
    \begin{overpic}
        [width=0.3\linewidth]{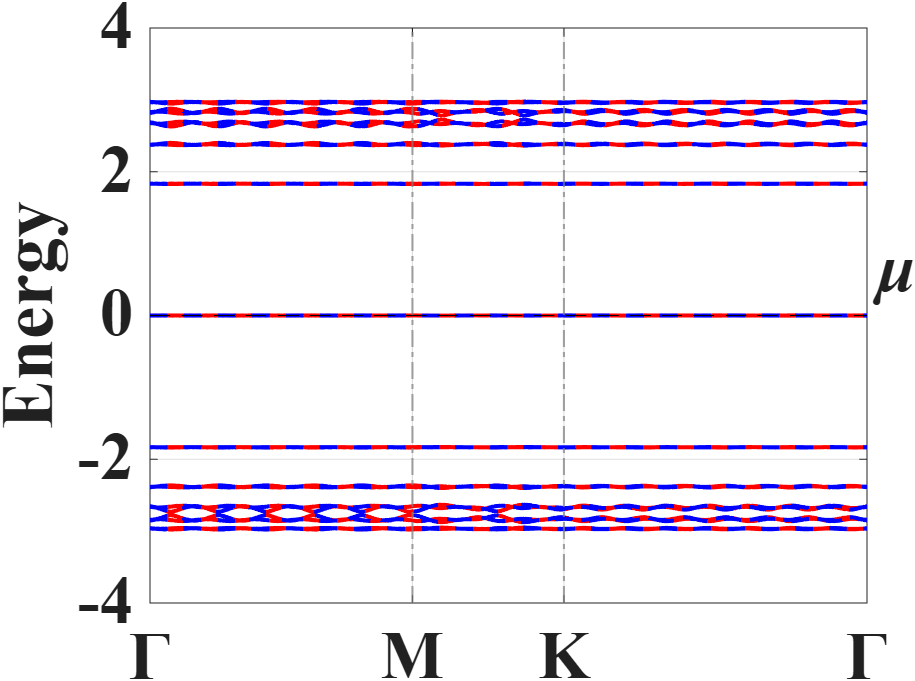}
        \put(0,70){(b)}
    \end{overpic}
    \begin{overpic}
        [width=0.3\linewidth]{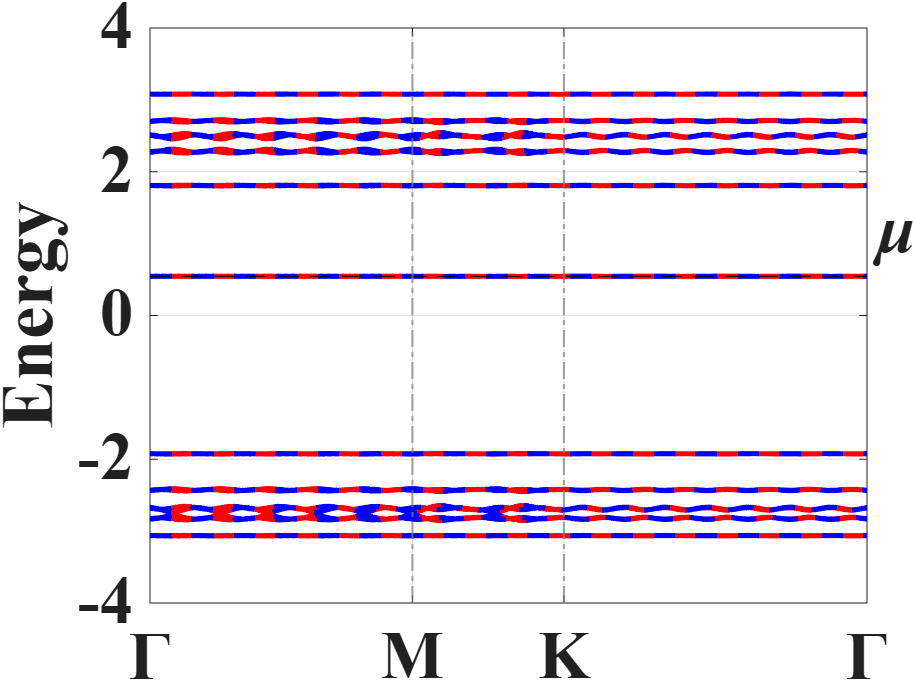}
        \put(0,70){(c)}
    \end{overpic}
    \\
    \caption{
        (a) Combined flux pattern on the triangular lattice.
        For uniform gauge flux $\phi_u=2\pi p/(4q)$ threading through the triangular plaquettes, we define the Peierls phases as $\varphi_1(l,s)=\phi_u[2(2(l-1)+s)-1/2]$ and $\varphi_2(l,s)=\phi_u[2(2(l-1)+s)+1/2]$, with sublattice index $s=1,2$.
        The magnetic unit cell is enlarged and contains $2q$ sites due to magnetic translational symmetry. 
        For example, the cyan parallelogram shows the enlarged magnetic unit cell for flux $\phi_u = 2\pi/(4q)$ with $q = 2$.
        Within each unit cell, sublattice sites are labeled as 1 (even) and 2 (odd) according to the spin liquid flux pattern.
        Spinon band structures for uniform flux $\phi_u = 2\pi/(4\times11)$ with staggered flux $\theta=0$ (panel b) and $\theta=\pi/6$ (panel c), respectively.
        For $\theta=0$ (panel b) the original two Chern bands split into 11 pairs of degenerate subbands with Chern numbers $[-2,\,9,\,-2,\,-2,\,-2,\,-2,\,-2,\,-2,\,-2,\,9,\,-2]$, while for $\theta=\pi/6$ (panel c) the sequence becomes $[-2,\,9,\,-2,\,-2,\,-2,\,-2,\,-2,\,-2,\,9,\,-2,\,-2]$.
    }
    \label{fig:Trianflux-theta-uphi}
\end{figure}

Additionally, considering the spin liquid two-sublattice flux pattern illustrated in Fig.~\ref{fig:Trianflux-theta-uphi}(a), the corresponding Hamiltonian can be expressed as 
$H(\theta,\phi_u)=-t\sum_{\bm{k}} 
\mathcal{H}_{\bm{k}}^{(1)}+\mathcal{H}_{\bm{k}}^{(2)}+\mathcal{H}_{\bm{k}}^{(3)}+h.c.$ with
\begin{equation}
\mathcal{H}_{\boldsymbol{k}}^{(1)}=
\sum_{l=1}^{q_{db}}
f^\dagger_{l,1}f_{l,1}
e^{i\bm{k}\cdot\bm{a}_2} 
e^{-i\varphi_1(l,1)}
+
f^\dagger_{l,2}f_{l,2}e^{i\bm{k}\cdot\bm{a}_2}
e^{i\pi} e^{-i\varphi_1(l,2)}
\end{equation}
\begin{equation}
\begin{aligned}
\mathcal{H}_{\boldsymbol{k}}^{(2)}=
\sum_{l=1}^{q_{db}}
f^\dagger_{l,1}f_{l,2} e^{i\theta}
+
\sum_{l=1}^{q_{db}-1}
f^\dagger_{l,2}f_{l+1,1} e^{i\theta}
+
f^\dagger_{q_{db},2}f_{1,1}
e^{i\bm{k}\cdot q_{db}\cdot2\bm{a}_1} e^{i\theta}
\end{aligned}
\end{equation}
\begin{equation}
\begin{aligned}
\mathcal{H}_{\boldsymbol{k}}^{(3)}&=
f^\dagger_{1,1} f_{q_{db},2}
e^{i\bm{k}\cdot(-q_{db}\cdot2\bm{a}_1+\bm{a}_2)} e^{-i\pi/2}
e^{-i\varphi_2(l=1,1)}
+
\sum_{l=2}^{q_{db}}
f^\dagger_{l,1} f_{l-1,2}
e^{i\bm{k}\cdot\bm{a}_2} e^{-i\pi/2}
e^{-i\varphi_2(l,1)}
\\&+
\sum_{l=1}^{q_{db}}
f^\dagger_{l,2}f_{l,1} e^{i\bm{k}\cdot\bm{a}_2} e^{i\pi/2}
e^{-i\varphi_2(l,2)}
\end{aligned}
\end{equation}
where $\varphi_1(l,s)=\phi_u[2\cdot(2(l-1)+s)-1/2]$, and 
$\varphi_2(l,s)=\phi_u[2\cdot(2(l-1)+s)+1/2]$, with $\phi_u = 2\pi\,p/(4q)$ and sublattice index
$s=1,2$.
In this combined flux effect, the enlarged magnetic unit cell has instead $2q$ sites.
The energy spectrum is obtained by numerically diagonalizing the $2q \times 2q$ Hamiltonian matrix for each wave vector ${\bm{k}}$.
Therefore, the original Chern bands split into several magnetic subbands with high flatness and different Chern numbers as shown in Fig.~\ref{fig:Trianflux-theta-uphi}(b,c).

\subsection{Real-space torus gauge}
\label{sm:RealSpaceGauge}
While continuum magnetic fields are often implemented using Landau or symmetric gauges, such choices are ill-suited for finite lattice systems with periodic boundary conditions. To describe a situation in which only a single flux quantum threads the entire system in real space-a setup essential for Monte Carlo simulations-we instead adopt a real-space torus gauge. In this construction, the gauge field is defined directly on lattice bonds of an $L_x \times L_y$ torus, ensuring a uniform flux $\Phi_u$ through every triangular plaquette while maintaining exact periodicity.

However, to consider a scenario where only a single flux quantum penetrates the entire lattice system in real space \cite{wietek2024QuantumElectrodynamicsDimensions}, a different gauge choice is required---one that is essential for Monte Carlo calculations.
We construct a gauge field $A_{ij}$ on a torus ($L_x \times L_y$) to generate a uniform flux $\Phi_u$ per triangular plaquette. The bond phase $\theta_{ij}$ for a hopping from site $i=(n_1,n_2)$ to $j$ depends on the bond direction $\mathbf{d}$ and boundary conditions. We define phase increments $\delta_1 = 4\Phi_u$ and $\delta_2 = 2\Phi_u$.

\textbf{1. Horizontal Bonds ($\mathbf{d} = \mathbf{a}_1$)}
\begin{equation}
\theta(\mathbf{r}, \mathbf{a}_1) = 
\begin{cases} 
(n_1 - 1)\delta_1 + (n_2 - 1)\delta_2 & \text{Bulk} \\
(L_x - 1)\delta_1 - (n_2 - 1)((L_x - 1)\delta_1 + 2\Phi_u) & \text{Boundary } (n_1 = L_x)
\end{cases}
\end{equation}

\textbf{2. Vertical Bonds ($\mathbf{d} = \mathbf{a}_2$)}
\begin{equation}
\theta(\mathbf{r}, \mathbf{a}_2) = 
\begin{cases} 
(n_2 - 1)\delta_2 + (n_1 - 1)\delta_1 - \Phi_u & \text{Bulk} \\
((L_y - 1)\delta_2 - \Phi_u) - (n_1 - 1)((L_y - 1)\delta_2 - 2\Phi_u) & \text{Boundary } (n_2 = L_y)
\end{cases}
\end{equation}

\textbf{3. Diagonal Bonds ($\mathbf{d} = \mathbf{a}_2 - \mathbf{a}_1$)}
\begin{equation}
\theta(\mathbf{r}, \mathbf{a}_3) = 
\begin{cases} 
0 & \text{Bulk} \\
(n_2 - 1)(\theta_{\mathbf{a}_2}(L_x, 2) + 3\Phi_u) & \text{Left Edge } (n_1=1, 1 < n_2 < L_y) \\
-(n_1 - 2)(\theta_{\mathbf{a}_1}(2, L_y-1) - 2\Phi_u) & \text{Top Edge } (n_2=L_y, n_1 > 1) \\
\theta_{\mathbf{a}_2}(1, L_y) - \theta_{\mathbf{a}_1}(L_x, 1) - \Phi_u & \text{Top-Left Corner } (n_1=1, n_2=L_y)
\end{cases}
\end{equation}
This gauge choice ensures uniform flux $\Phi_u$ through all plaquettes on the torus.

\begin{figure}[t]
    \begin{overpic}
        [width=0.3\linewidth]{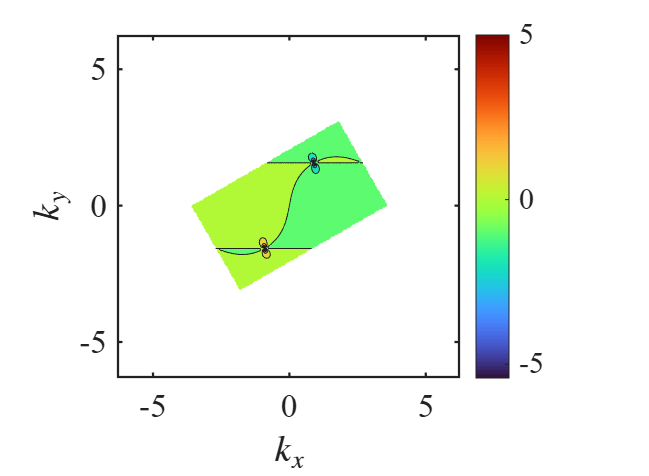}
        \put(0,65){(a)}
    \end{overpic}
    \begin{overpic}
        [width=0.3\linewidth]{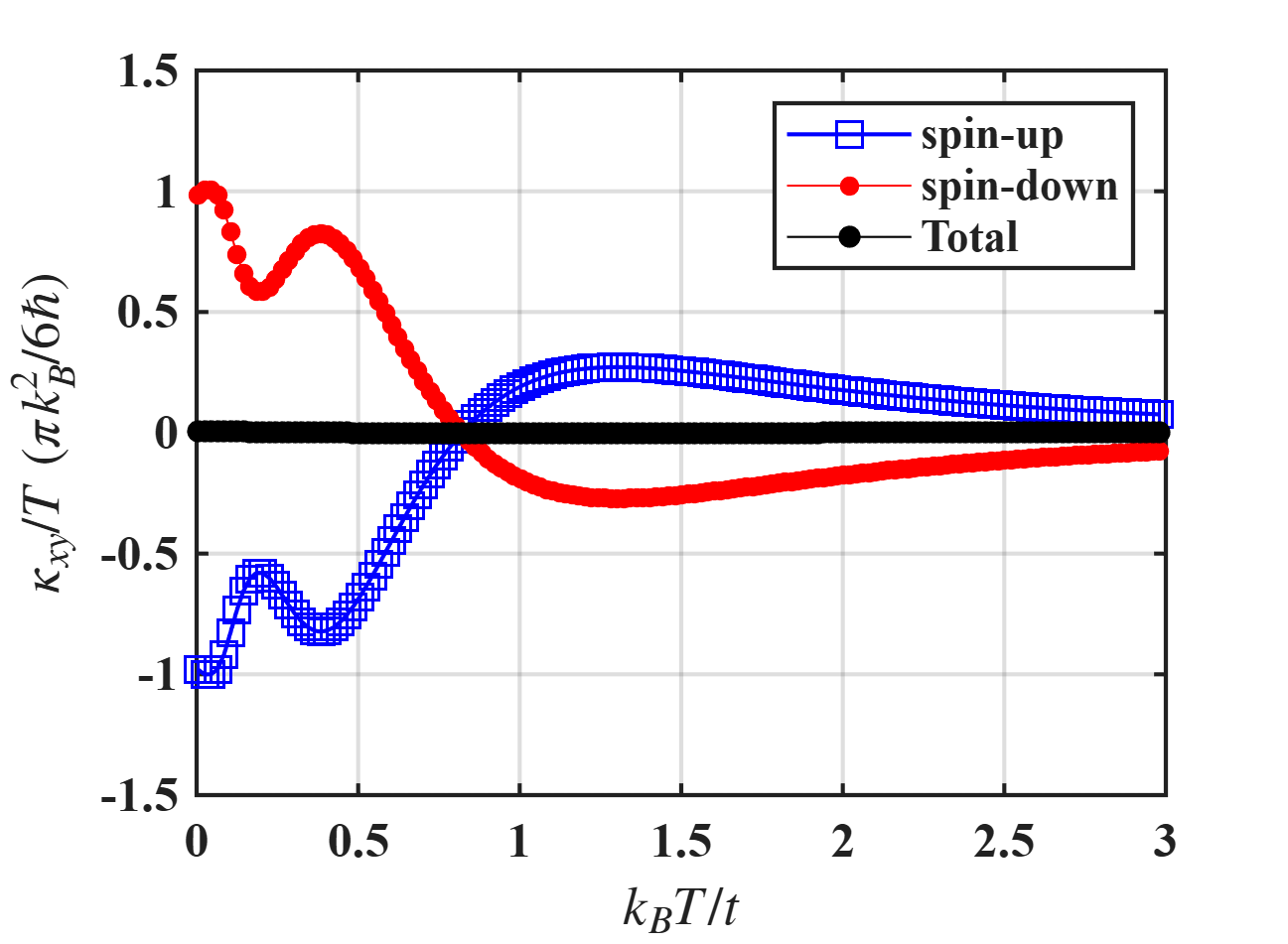}
        \put(0,65){(b)}
    \end{overpic}
        \begin{overpic}
        [width=0.3\linewidth]{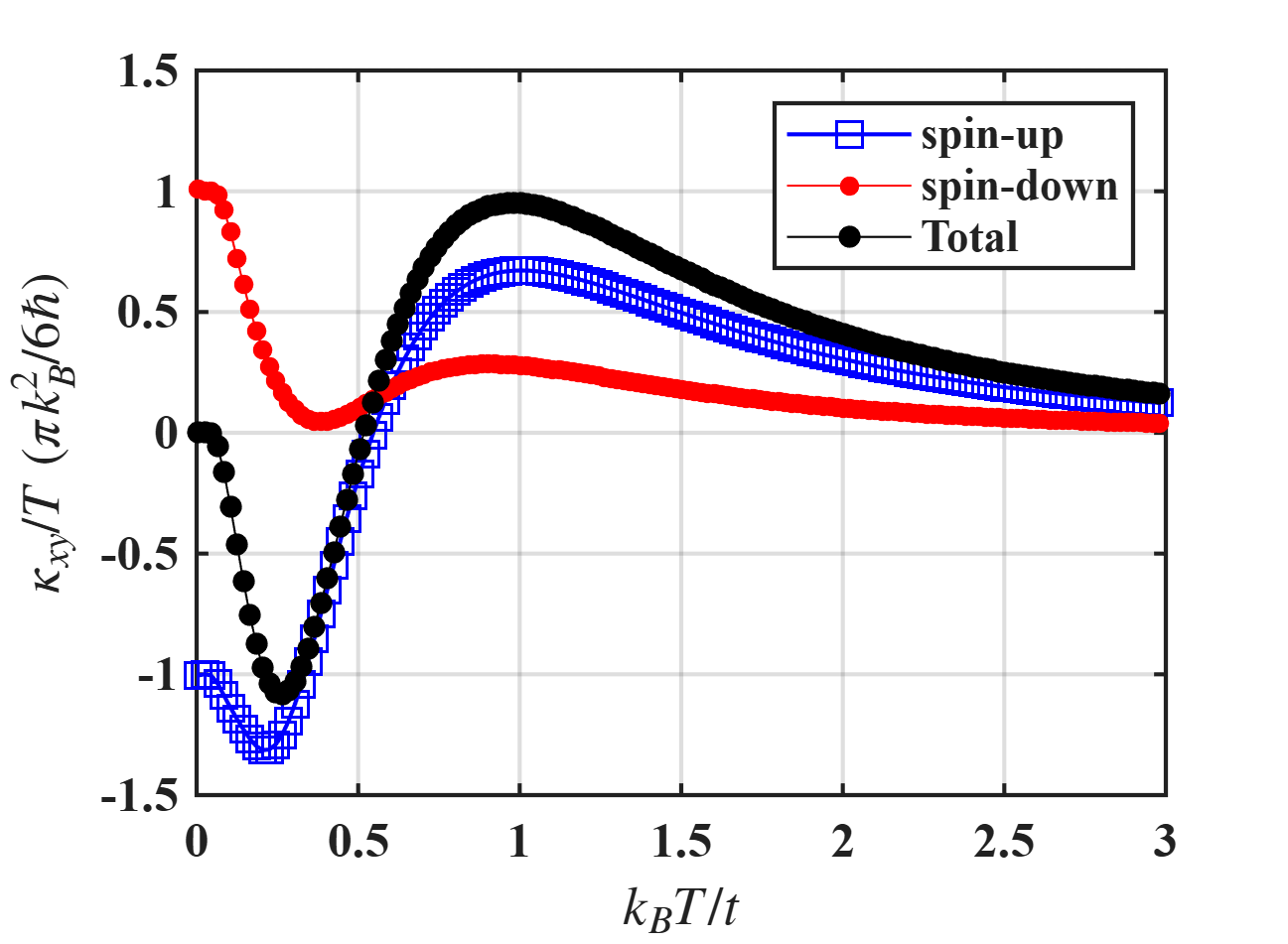}
        \put(0,70){(c)}
    \end{overpic}
    \caption{(a) Berry curvature distribution of the Dirac spin liquid 
    with uniform flux $\phi_u=0$ and Zeeman coupling $B=0$. 
    (b) Thermal Hall response with uniform flux $\phi_u=2\pi\,/76$, 
    staggered flux $\theta=0$ and $B=1$.
    (c) Similar to (b), but with $\theta=\pi/6$.
    }
    \label{fig:BerryCurv}
\end{figure}

\begin{figure}[t]
    \centering
    \begin{overpic}
        [width=0.3\linewidth]{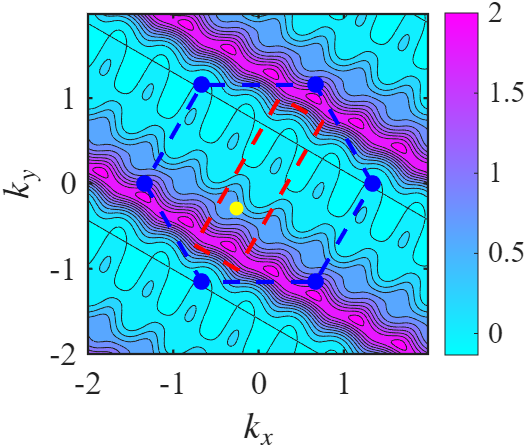}
        \put(0,65){(a)}
    \end{overpic}
    \begin{overpic}
        [width=0.3\linewidth]{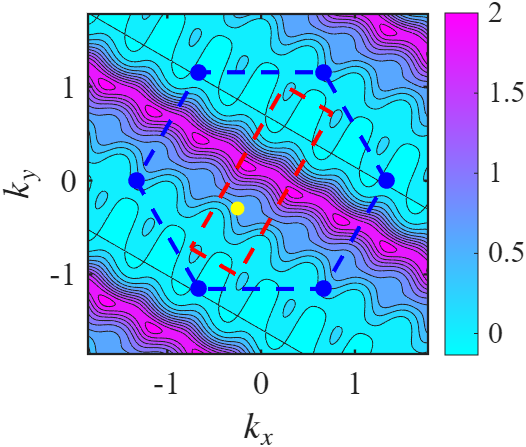}
    \end{overpic}
    \caption{Berry curvature distribution of the first (a) and fourth (b) magnetic subbands
    in the LL state with uniform flux $\phi_u=2\pi/8$
    The Chern numbers of these two subbands both are $C=1$. 
    And the yellow points denote the shifted inversion centers in the magnetic Brillouin zone.
    }
    \label{fig:BerryCurv_sym}
\end{figure}

\section{More discussion on thermal Hall effect}
\label{app:DSL}

The absence of a finite thermal Hall response in a U(1) Dirac spin liquid 
is rooted in the symmetry-constrained structure of the spinon Berry curvature.
In the Dirac spin liquid, the low-energy spinon spectrum consists of a pair of massless Dirac cones located at symmetry-related momenta in the Brillouin zone. Although each Dirac point carries a quantized Berry phase, 
the associated Berry curvature is singular and localized at the Dirac nodes, vanishing elsewhere in
the momentum space. As a result, the Berry curvature contributions from different Dirac cones cancel exactly 
upon Brillouin-zone integration.
Moreover, time-reversal symmetry enforces an odd transformation of the Berry curvature, 
$\Omega(\mathbf{k}) = -\Omega(-\mathbf{k})$ in Fig.~\ref{fig:BerryCurv}(a), 
while the spinon occupation remains uniform in momentum space. Consequently, 
the Berry-curvature-weighted integral governing intrinsic Hall-type responses vanishes identically. 
It is worth mentioning that DSL Hamiltonian has symmetric spectrum due to particle-hole symmetry, i.e. $P H(\mathbf{k}) P^{-1} = -H(-\mathbf{k})$ with $P=i\sigma_y K$.
This symmetry ensures that for every state at energy $\epsilon$  there exists a corresponding state at energy $-\epsilon$.

Now we first turn to the Fermi-pocket state with a finite Zeeman coupling.
Upon introducing a Zeeman coupling $B$ that breaks time-reversal, 
the spinon bands split into the Fermi-pocket state, 
but the overall symmetric Berry curvature remain unchanged and each bands are equally 
filled under the chemical potential. Each spin sector still hosts Dirac cones 
with opposite Berry curvature distributions, 
leading to a continued cancellation in the total thermal Hall conductivity. 
Thus, even with a finite $B$, the Fermi-pocket state maintains a zero thermal Hall 
response due to the fundamental symmetry constraints on its Berry curvature structure.

Next we discuss the LL state after including the spontaneous flux generation. With 
a uniform flux $\phi_u$ and Zeeman coupling $B$, the spinon bands in the Dirac spin liquid split into multiple Hofstadter-like subbands for LL state, each characterized by distinct Chern numbers in Fig.~\ref{fig:Trianflux-theta-uphi}(b).
Although each spin sector individually acquires a finite thermal Hall response (blue and red lines) 
in Fig.~\ref{fig:BerryCurv}(b), the total thermal Hall conductivity (black line) vanishes 
at all temperatures.
This can be understood once the Hall coefficient $\sigma_{xy}$ is analyzed to be an odd function of energy.
By defination, 
\begin{equation}
    \begin{split}
    \sigma_{xy}(\epsilon)
    &=
    -\sum_n\sum_{\epsilon_{n\boldsymbol{k}}<\epsilon}
    \Omega_{n\boldsymbol{k}}(\epsilon_{n\boldsymbol{k}})
    =
    -\int_{-\infty}^{\epsilon}d\gamma
    \int_{\epsilon_{\boldsymbol{k}}=\gamma}\frac{d^2\boldsymbol{k}}{(2\pi)^2}\,
    \Omega_{\boldsymbol{k}}(\epsilon_{\boldsymbol{k}}),
    \\
    \sigma_{xy}(-\epsilon)&=
    -\int_{-\infty}^{-\epsilon}d\gamma
    \int_{\epsilon_{\boldsymbol{k}}=\gamma}\frac{d^2\boldsymbol{k}}{(2\pi)^2}\,
    \Omega_{\boldsymbol{k}}(\epsilon_{\boldsymbol{k}})
    =
    -\int_{\epsilon}^{\infty}d\gamma
    \int_{\epsilon_{\boldsymbol{k}}=-\gamma}\frac{d^2\boldsymbol{k}}{(2\pi)^2}\,
    \Omega_{\boldsymbol{k}}(\epsilon_{\boldsymbol{k}}).
    \label{eq:sigmaxy}
    \end{split}
\end{equation}
where, for simplicity, we absorb the subband index $n$ into the continuous energy variable $\gamma$.
For the LL state derived from the DSL, we numerically find that for every $\Omega_{\boldsymbol{k}}(\epsilon_{\boldsymbol{k}})$ with energy $\epsilon_{\boldsymbol{k}}=\gamma$,
there exists a corresponding $\Omega_{\bar{\boldsymbol{k}}}(\epsilon_{\bar{\boldsymbol{k}}})=\Omega_{\boldsymbol{k}}(\epsilon_{\boldsymbol{k}})$ with energy $\epsilon_{\bar{\boldsymbol{k}}}=-\gamma$, leading to the equavalant contribution for the momentum integral.
And the momenta $\boldsymbol{k}$ and $\bar{\boldsymbol{k}}$ can be related by a shifted inversion center (yellow points) in the Brillouin zone in the~\ref{fig:BerryCurv_sym}.
This special property should be related to the underlying gauge symmetries realized in the LL state,
which we defer for future study.
Therefore one can prove $\sigma_{xy}(\epsilon)$ is an odd function of energy $\epsilon$ from the addition of two equations in Eq.~\eqref{eq:sigmaxy} and the fact that the total Berry curvature, i.e. Chern number, is zero.
Under finite Zeeman splitting, while the individual contributions $\sigma_{xy}^{\pm}\equiv \sigma_{xy}(\epsilon\pm E_{Zeeman})$ for spin-$\uparrow$ and spin-$\downarrow$ spinons are neither strictly even nor odd, their sum $\sigma_{xy}^{\text{tot}}=\sigma_{xy}^{+} + \sigma_{xy}^{-}$ remains an odd function of energy.
Consequently, when integrated over energy with the remaining even-function factors in Eq.~(4), the total thermal Hall conductivity vanishes identically at any temperature.

While the vanishing thermal Hall conductivity in the LL state can be attributed to the underlying symmetries of this special LL state, this cancellation is not generic for all flux states.
Especially, the chiral spin liquid phase breaks time-reversal symmetry intrinsically, 
leading to an asymmetric distribution of Berry curvature among the Hofstadter subbands. 
This asymmetry allows for a finite thermal Hall conductivity, 
as shown in Fig.~\ref{fig:BerryCurv}(c).
And also, further neighbor hopping terms may break the special symmetry in LL state,
resulting in a finite thermal Hall conductivity.

\section{Monte Carlo method for projected wavefunction}
\label{app:MC}

To determine the ground state energy and characterize potential spin ordering 
under Gutzwiller projection, we employ the Monte Carlo sampling \cite{becca2017QuantumMonte,gros1989PhysicsProjected}. 
The approach is based on evaluating the expectation value of an operator $\hat{O}$ 
with respect to a given wavefunction $\ket{\Psi}$:
\begin{equation}
  \expval*{\hat{O}} = \frac{\mel{\Psi}{\hat{O}}{\Psi}}{\braket{\Psi}{\Psi}} 
  = \frac{\sum_x \braket{\Psi}{x} \mel{x}{\hat{O}}{\Psi}}{\sum_x \braket{\Psi}{x}\braket{x}{\Psi}} ,
\end{equation}
where $\ket{x}$ is a complete set of basis. Theoretically, if we can enumerate all the basis states $\ket{x}$, we can calculate the expectation value exactly. However, in practice, the Hilbert space is usually exponetially large with the number of particles, making the exact calculation infeasible. Therefore, we resort to Monte Carlo method to carry out importance sampling, and the key step is to rewrite the expectation value as
\begin{equation}
  \expval*{\hat{O}} = \sum_x P(x) O_L(x) , \quad P(x) = \frac{\abs{\braket{x}{\Psi}}^2}{\sum_{x'} \abs{\braket{x'}{\Psi}}^2} ,
\end{equation}
where we introduce the local estimator of the operator $\hat{O}$:
\begin{equation}
  O_L(x) = \frac{\mel{x}{\hat{O}}{\Psi}}{\braket{x}{\Psi}} = \sum_{x'} \mel{x}{\hat{O}}{x'} \frac{\braket{x'}{\Psi}}{\braket{x}{\Psi}} .
\end{equation}
It is straightforward to verify that $P(x)$ is a normalized probability distribution. We can therefore sample the basis states $\ket{x}$ according to $P(x)$ using the Markov Chain Monte Carlo (MCMC) method, generating a sequence of samples $\{x_i\}$ that follow this distribution. 
The expectation value is then calculated as the average of the local estimator:
\begin{equation}
  \expval*{\hat{O}} \approx \frac{1}{N} \sum_{i=1}^N O_L(x_i) .
\end{equation}

\begin{figure}
    \centering
    \includegraphics[width=0.35\textwidth]{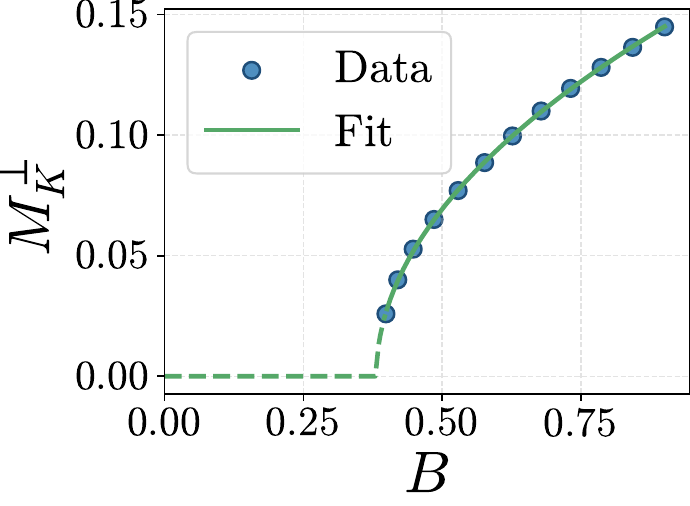}
    \caption{The transverse magnetization $M_K^\perp$ as a function of Zeeman coupling $B$ for the projected wavefunction on an $18\times18$ lattice with $\theta=0.1\pi$, $J_\chi=0.3$. The fitting curve is $M_K^\perp = 0.201 (B-0.381)^{0.5}$.}
    \label{fig:MK_CSL}
\end{figure}

In our studied case, we use the Gutzwiller projected mean-field wavefunction as the trial wavefunction $\ket{\Psi}$, which is constructed from a mean-field Hamiltonian that captures the essential physics of the system. 
More specifically, we start from a mean-field Hamiltonian, and obtain its ground state wavefunction $\ket{\Psi_{\text{MF}}}$ via diagonalizing. Finally, we apply the Gutzwiller projection operator $P = \prod_i (1 - n_{i \uparrow} n_{i \downarrow})$ to enforce the single-occupancy constraint, resulting in the wavefunction $\ket{\Psi} = P \ket{\Psi_{\text{MF}}}$.
To incorporate the uniform flux in Monte Carlo calculations, we adopt the gauge construction detailed in Sec.~\ref{sm:RealSpaceGauge}. 
This gauge choice ensures that the uniform flux $\phi_u$ threading the lattice is self-consistently related to the finite magnetization, i.e., $\phi_u = (2\pi) m_z / 2$, which is also consistent with the continuum theory analysis.
Additionally, we put the results of $M_K^\perp$ vs $B$ in Fig.~\ref{fig:MK_CSL}, which is obtained from the projected wavefunction calculation on an $18\times18$ lattice with $\theta=0.1\pi$, $J_\chi=0.3$.

\end{document}